\documentclass[journal]{IEEEtran}

\usepackage{xcolor,soul,framed} 

\colorlet{shadecolor}{yellow}
\usepackage[pdftex]{graphicx}
\graphicspath{{../pdf/}{../jpeg/}}
\DeclareGraphicsExtensions{.pdf,.jpeg,.png}

\usepackage{array}
\usepackage{mdwmath}
\usepackage{mdwtab}
\usepackage{eqparbox}
\usepackage{cite}
\usepackage{amsmath,amssymb,amsfonts,dsfont}
\usepackage{algorithmic}
\usepackage[pdftex]{graphicx}
\usepackage{textcomp}
\usepackage{caption}
\usepackage{color,soul}
\usepackage{url}
\usepackage{subfigure}
\usepackage{epstopdf}
\usepackage{multirow}
\usepackage{bm}
\usepackage{ulem}
\begin{document}
\bstctlcite{IEEEexample:BSTcontrol}
    \title{Interlayer Link Prediction in Multiplex Social Networks Based on Multiple Types of Consistency between Embedding Vectors}
  \author{Rui Tang, Zhenxiong Miao, Shuyu Jiang, Xingshu Chen, Haizhou Wang, and Wei Wang

\thanks{This work was supported by the National Natural Science Foundation of China under Grant Nos. U19A2081, 81602935, 81773548, 61802270, and 61802271; the
Science and Engineering Connotation Development Project of Sichuan University under Grant No. 2020SCUNG129; the Sichuan Science and Technology Program under Grant No. 20YYJC4001. (Corresponding author:Xingshu Chen, Haizhou Wang. e-mail: chenxsh@scu.edu.cn, whzh.nc@scu.edu.cn)}
  \thanks{Rui Tang, Zhenxiong Miao, and Shuyu Jiang are with School of Cyber Science and Engineering, Sichuan University, Chengdu 610065, China.}
  \thanks{Xingshu Chen and Haizhou Wang are with School of Cyber Science and Engineering, Sichuan University, Chengdu 610065, China; Cyber Science Research Institute, Sichuan University, Chengdu 610065, China.}%
  \thanks{Wei Wang are with Cyber Science Research Institute, Sichuan University, Chengdu 610065, China.}}

\markboth{
}{Rui Tang \MakeLowercase{\textit{Tang et al.}}: Interlayer Link Prediction in Multiplex Social Networks Based on Multiple Types of Consistency between Embedding Vectors}

\maketitle

\begin{abstract}
Online users are typically active on multiple social media networks (SMNs), which constitute a multiplex social network. With improvements in cybersecurity awareness, users increasingly choose different usernames and provide different profiles on different SMNs. Thus, it is becoming increasingly challenging to determine whether given accounts on different SMNs belong to the same user; this can be expressed as an interlayer link prediction problem in a multiplex network. To address the challenge of predicting interlayer links , feature or structure information is leveraged. Existing methods that use network embedding techniques to address this problem focus on learning a mapping function to unify all nodes into a common latent representation space for prediction; positional relationships between unmatched nodes and their common matched neighbors (CMNs) are not utilized. Furthermore, the layers are often modeled as unweighted graphs, ignoring the strengths of the relationships between nodes. To address these limitations, we propose a framework based on multiple types of consistency between embedding vectors (MulCEV). In MulCEV, the traditional embedding-based method is applied to obtain the degree of consistency between the vectors representing the unmatched nodes, and a proposed distance consistency index based on the positions of nodes in each latent space provides additional clues for prediction. By associating these two types of consistency, the effective information in the latent spaces is fully utilized. Additionally, MulCEV models the layers as weighted graphs to obtain representation. In this way, the higher the strength of the relationship between nodes, the more similar their embedding vectors in the latent representation space will be. The results of our experiments on several real-world and synthetic datasets demonstrate that the proposed MulCEV framework markedly outperforms current embedding-based methods, especially when the number of training iterations is small.
\end{abstract}


\IEEEpeerreviewmaketitle
	
\section{Introduction}
Social media network (SMN) applications have significantly enriched the daily lives of users and have attracted the attention of many researchers~\cite{centola2010spread,ShiChuan2017,ShuKai2017,wang2020efficient}. 
Online users often make use of several SMNs simultaneously, for example by recording individual impressions of current events on Twitter, sharing photographs on Instagram, and searching for job information on LinkedIn. These various SMNs thus form a multiplex social network~\cite{nguyen2013least,jiang2013understanding,Kivela2014,dadlani2017mean,liu2019infection,zhang2018optimizing}, of which each SMN constitutes a layer. Accounts are represented as nodes, and friendship relations or interactions are represented as intralayer links. If two accounts in different SMNs belong to the same user, an interlayer link exists between the corresponding nodes across different layers. The structure of a multiplex network has a significant influence on cascades~\cite{Gao2011a}, propagation~\cite{Wangwei2019,xu2018spectral}, synchronization~\cite{liu2020intralayer,zhao2018pinning}, and games~\cite{santos2008social}.

\begin{figure*} [t!]
    \centering
    \includegraphics[width=0.99\textwidth]{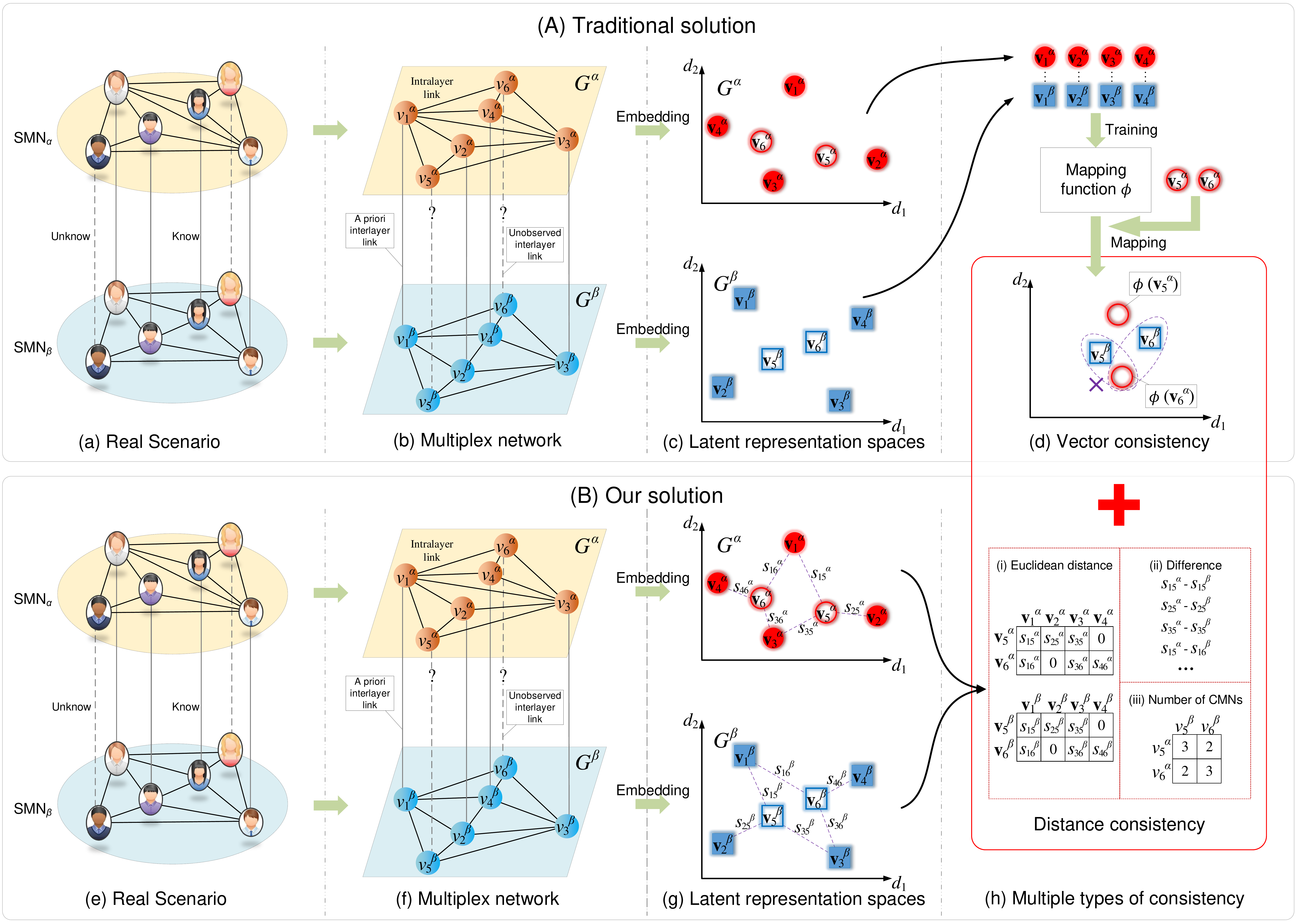}
    \caption{Example of interlayer link prediction based on multiple types of consistency between embedding vectors. (A) Traditional solution of using network embedding techniques for interlayer link prediction. (a) Real scenario: There are two SMNs, each of which has six user accounts. The black lines indicate the friendship relations or interactions of the accounts. The accounts linked by the vertical gray line belong to the same user. The correspondence of the account pairs linked by a solid gray line is known in advance; the task is to determine the correspondence of the other accounts. (b) Multiplex network: The multiple SMNs are represented by a multiplex network. The four vertical solid gray lines are represented as the a priori interlayer links. The task 
    becomes the task of predicting the unobserved interlayer links in the multiplex network. (c) Latent representation spaces: Network embedding techniques are used to address the interlayer link prediction problem. Each layer of the multiplex network is embedded into a latent representation space. 
	(d) Vector consistency. The nodes with a priori interlayer link are used to train the mapping function to unify nodes into a common latent space. The unobserved interlayer links are predicted by vector consistency. (B) Our solution. (e) and (f) are the same with (a) and (b), respectively. (g) Latent representation spaces: The Euclidean distance between each unmatched node and its matched neighbors can be calculated. (h) Multiple types of consistency: The degree of match to estimate whether an interlayer link exists between two unmatched nodes across different layers consists of two parts. One is the degree of vector consistency obtained by the traditional solution, and the other is the degree of distance consistency.
}
    \label{pic:motivation}
\end{figure*}

The goal of interlayer link prediction is to leverage feature or structure information to determine whether accounts across different SMNs belong to the same user~\cite{tang2020interlayer}; this is a challenging task in multiplex network analysis. It is also known as anchor link prediction~\cite{KongXiangnan2013,cheng2019deep}, network alignment~\cite{zhang2019multilevel,ChuXiaokai2019-www,li2019adversarial,zhou2019translink,zhou2019disentangled}, user identification~\cite{zafarani2015user}, and user identity linkage~\cite{zhou2019translink}.

As hacking attempts have become more frequent, online users' security awareness has gradually increased. An increasing number of users create accounts under different usernames, hide profile information, or even provide fake content on profile configuration pages~\cite{li2020exploiting}. In this way, users access SMNs anonymously to make friends, share information, and discuss problems, thereby not only using multiple SMNs simultaneously but also protecting their privacy. However, such anonymity can pose a certain degree of harm to society. Cybercriminals register a large number of accounts on multiple SMNs and subsequently engage in various types of illegal activities. For example, they might circulate messages containing untruths, spread malware program links, or initiate financial fraud on these SMNs~\cite{HanXiaohui2017}. Predicting the interlayer links of the multiplex network comprising different SMNs can help criminal investigation authorities establish cybercriminals' patterns of law violations, model their online behaviors, identify their regions of activity, or even determine their real-world identities, thereby effectively fighting cybercrime.

There are other benefits as well to predicting the interlayer links. For instance, because information and rumors typically spread across multiple SMNs, predicting interlayer links can help improve the understanding of information diffusion~\cite{li2015multiple}. Furthermore, the use of SMNs as evidence in trials related to issues of custody, divorce, and insurance is rising rapidly~\cite{arshad2019evidence}. A method for identifying the interlayer links across multiple SMNs would be a powerful tool in the collection of evidence for civil and criminal investigations.

The problem of predicting interlayer links is typically solved by leveraging feature or structure information accessed from the multiple SMNs. Currently, there are three main approaches for handling this problem: (i) feature-based prediction~\cite{ZhaoDongsheng2018,Zafarani2009,Perito2011,Zafarani2013,LiuJing2013-WSDM,li2019matching,Carmagnola2009,Iofciu2011,AbelFabian2013,Goga2013,Cortis2013, MuXin-KDD2016,fu2020deep,Riederer2016,chen2018effective,feng2019dplink,ZhengRong2006,narayanan2012feasibility,Goga2013a}, (ii) network-based prediction~\cite{Narayanan2009,korula2014efficient,ZhouXiaoping2016,tang2020interlayer,ren2019meta,ZhuYuanyuan2012,zafarani2015user,ZhangYutao2015,zhang2015multiple,ZhangJiawei2015-IJCAI,ZhangSi2016-KDD,ZhangSi2019-www,TanShulong2014-AAAI,LiuLi2016,liu2019structural,ManTong2016-IJCAI,ZhouFan2018,zhou2019disentangled,ZhouXiaoping2018-IEEE,WangYongqing2019-www,cheng2019deep,li2019adversarial,ChuXiaokai2019-www,wang2019user,wang2019anchor,zhou2019translink}, and (iii) a combination of multiple approaches~\cite{JainParidhi2013,Nunes2012,KongXiangnan2013}. Of these, network-based methods have attracted more attention; they show increasing promise because few people share the same circle of friends~\cite{ZhouXiaoping2016}, and information on connections in SMNs is quite easy to obtain~\cite{fu2020deep}.

With the development of neural networks and deep learning~\cite{chen2020random,chen2015fuzzy}, network embedding techniques~\cite{wang2020node} have been utilized to learn latent, low-dimensional representations of network nodes while preserving network structure. After all of the nodes are represented as low-dimensional vectors into the latent representation space, advanced network analytic tasks such as node classification, community detection, and link prediction can be efficiently carried out~\cite{zhang2018network}. Motivated by the advances in network embedding techniques for single-network tasks, researchers have proposed several strategies to leverage these techniques for solving the interlayer link prediction problem~\cite{ManTong2016-IJCAI,LiuLi2016,ZhouFan2018}. Typically in these studies, network embedding techniques are used first, to learn the latent representations of nodes in different layers of the multiplex network. After that, a priori interlayer node pairs are constrained to have the same latent representations to unify nodes into a common latent representation space. Finally, the unobserved interlayer links are predicted by comparing the embedding vectors of unmatched nodes across different layers in the common latent space.

To unify all nodes into a common latent space and predict the unobserved interlayer links, most embedding-based methods utilize a priori interlayer links to train an approximate mapping function after achieving the latent representations, as was done in Refs.~\cite{MuXin-KDD2016,ManTong2016-IJCAI,ZhouFan2018,WangYongqing2019-www}. However, the perfect mapping function is difficult to obtain, as each layer's latent space is unknown to the others~\cite{ZhouFan2018} and the sampled contexts may differ when nodes are embedded into the latent representation spaces. This leads to unsatisfactory performance, especially when the number of training iterations is small. Apart from predicting the unobserved interlayer links by comparing the embedding vectors in the common latent space unified by the mapping function, the positional relationships between unmatched nodes and their common matched neighbors (CMNs) can also be used to measure whether an interlayer link exists between two unmatched nodes that lie in different layers. In other words, the effective information in the latent representation spaces is not fully utilized.

Figure~\ref{pic:motivation} illustrates the general components of interlayer link prediction based on multiple types of consistency between embedding vectors. As shown in Fig.~\ref{pic:motivation} (d), if we use only the information in the common latent space unified by the mapping function, the node $v^\alpha_6$ will be matched with $\bm{\mathrm{v}}^\beta_5$ as their embedding vectors are more consistent. However, if we analyze the positional relationships between unmatched nodes and their CMNs simultaneously, we might uncover more clues for predicting the unobserved interlayer links. We propose a ``distance consistency'' index to measure this relationship. As shown in Fig.~\ref{pic:motivation} (h), three aspects are considered in the distance consistency index: (i) the Euclidean distance between the unmatched node and its matched neighbor, (ii) the difference between two Euclidean distances formed by unmatched nodes across different layers and their CMNs, and (iii) the number of CMNs.

When each layer of the multiplex network is embedded into a latent representation space, different layers are often modeled as unweighted graphs, and the strength of the relationships between nodes is often ignored. However, the intralayer links between nodes may have different relationship strengths. For example, if a boy has only one friend, the friendship between him and his friend is highly likely to be closer than one between individuals who have many friends. To distinguish among relationship strengths, the intralayer links between nodes should be weighted. To address this problem, we use a weighted-embedding method to embed each layer of the multiplex network in the form of weighted graphs based solely on the network structure.

In this study, we developed a framework based on multiple types of consistency between embedding vectors (MulCEV) for interlayer link prediction in a multiplex network; it focuses on making full use of information in the latent representation spaces for prediction and on modeling different layers as weighted graphs to obtain better representation. The main contributions can be summarized as follows:

\begin{itemize}
\item We propose a distance consistency index that is based on the positions of nodes in each latent representation space, which leverages CMNs of the unmatched nodes across different layers as references to provide additional clues for predicting interlayer links. The degree of match to estimate whether an interlayer link exists between two unmatched nodes across different layers consists of two parts: the degree of vector consistency, which applies the traditional embedding-based method to measure the consistency of the embedding vectors of the unmatched nodes in the common latent representation space, and the degree of distance consistency proposed above. The effective information in the latent representation spaces is fully utilized by associating these two types of consistency between embedding vectors.
\item We model each layer of the multiplex network as a weighted graph to obtain representation based solely on the network structure. Thus, the higher the strength of the relationship between nodes, the closer are their embedding vectors in the latent representation space.
\item In order to reduce the time complexity, we adopt the technique of matrix multiplication to optimize the process of calculating the distance consistency and vector consistency for all unmatched node pairs.
\item We test the effectiveness of the proposed MulCEV framework on four widely used real-world and three synthetic multiplex network datasets and report the results against those of state-of-the-art methods.
\end{itemize}


\section{Related Work}
This section introduces the embedding-based interlayer link prediction methods. The remaining types of methods are introduced in Appendix A.

Network embedding techniques aim to represent nodes in a network by low-dimensional vectors in a latent representation space so that advanced analytic tasks, such as node classification, community detection, and link prediction, can be conducted more efficiently in both time and space~\cite{cui2018survey}. DeepWalk~\cite{PerozziBryan2014} leverages uniform random walks to generate a set of node sequences that are similar to the word sequences in natural language and uses the skip-gram model to learn vertex representations of nodes. Node2vec~\cite{grover2016node2vec} demonstrated that DeepWalk is not sufficiently expressive to capture a more global structure and incorporated a biased random walk strategy to improve it. Tang et al.~\cite{TangJian2015} proposed a large-scale information network embedding (LINE) approach to preserve both the first- and second-order proximities between nodes. Subsequently, Wang et al.~\cite{wang2017community} preserved not only the first- and second-order proximity of vertices but also the community structure. There have been numerous other studies using network embedding techniques such as dynamic network embedding~\cite{zhiyuli2018modeling,du2018dynamic} and embedding for scale-free networks~\cite{FengRui2018}.

Increasingly, computer and network scientists are exploring ways of employing network embedding techniques to improve their ability to predict interlayer links in terms of accuracy, applicability, and efficiency.
Tan et al.~\cite{TanShulong2014-AAAI} tried to map accounts across SMNs based on network embedding, adopting hypergraphing to model high-order relations of SMNs and representing nodes in a common latent space. Their method infers correspondence by comparing distances between the vectors of the unmatched nodes. Liu et al.~\cite{LiuLi2016} represented the multiple SMNs with a shared latent space and determined the interlayer links by computing the cosine similarity between the latent space vector of one node in layer $\alpha$ and another in layer $\beta$. The network embedding process was integrated with the entity alignment process under a unified optimization framework.
They further refined their proposed method in Ref.~\cite{liu2019structural} by incorporating structural diversity. The structural diversity focuses on the impact of whether the a priori matched nodes come from different communities.

Instead of embedding all layers into a common latent space, Man et al.~\cite{ManTong2016-IJCAI} projected each SMN into a unique latent space and represented nodes by low-dimensional vectors in the latent space. Then, they trained a cross-layer mapping function for predicting interlayer links.
Zhou et al.~\cite{ZhouFan2018} adopted the same idea and proposed a semi-supervised approach that leverages dual learning to pretrain the mapping function to improve prediction accuracy. They focused on learning latent semantics of both the node representation and the network structure in Ref.~\cite{zhou2019disentangled}. Zhou et al.~\cite{ZhouXiaoping2018-IEEE} studied the scenario without a priori interlayer links and proposed an unsupervised approach for the prediction. Considering time complexity, Wang et al.~\cite{WangYongqing2019-www} proposed a framework that directly learns a binary hash code for each node across SMNs, which obtained high time efficiency while maintaining competitive prediction accuracy. In Ref.~\cite{cheng2019deep}, the authors adopted active learning to reduce the cost of labeling a priori node pairs. In Ref.~\cite{li2019adversarial}, the authors viewed all of the nodes in one layer as a whole and executed the prediction at the distribution level. Chu et al.~\cite{ChuXiaokai2019-www} considered multi-layer scenarios in which the number of layers is more than two. They refined two types of low-dimensional vectors for each node: an inter-vector for interlayer link prediction, and an intra-vector for downstream network analysis tasks.

In some studies, structural information and attribute information were embedded simultaneously to perform the interlayer link prediction. Wang et al.~\cite{wang2019user} proposed a linked heterogeneous network embedding (LHNE) method to fuse the content and structural information of a user into a unified latent representation space to identify account linkages. In Ref.~\cite{wang2019anchor}, the authors proposed a semi-supervised network embedding method to learn the attribute information and structural information simultaneously. Heterogeneous SMNs differ substantially in several aspects, including network structure, user behavior, and user information. TransLink~\cite{zhou2019translink} captures the heterogeneities of SMNs and embeds both nodes and their various types of interactions into a unified latent space.

In summary, most of the available embedding-based methods focus on optimizing the learning framework, reducing time consumed, predicting scenarios for more than two layers, embedding using multiple attributes, etc. These methods require mapping techniques to ensure that the embedding vectors of the correspondence nodes are equal irrespective of direct embedment of the layers into a unified latent space or mapping of separate spaces into a unified space. However, perfect mapping is difficult because the sampled contexts may differ when nodes are embedded into the latent representation spaces. Thus, the following problems should be explored: Is there an alternative method for prediction in lieu of only mapping and calculating vector similarities in the unified space? Can the positional relationships between unmatched nodes and their CMNs be used to determine if interlayer links exist between pairs of unmatched nodes in different layers? How can these relationships be leveraged for prediction? To address these, we developed a framework based on multiple types of consistency between embedding vectors for interlayer link prediction.
\section{Preliminaries and Problem Statement}
In this section, we define related terminologies and explain the problem of interlayer link prediction. The main symbols and notations are shown in Table III in Appendix B.

\subsection{Definitions}
In general, an SMN can be represented as a graph $G(V,E)$, where $V$ is a node set representing all the accounts, and $E \subseteq V \times V$ is an edge set representing the relationships among the accounts. Multiple SMNs can constitute a multiplex network.

\textbf{Definition 1: Multiplex network. }
Given a set of SMNs, we can denote them using superscripts $\alpha, \beta, \dots$, such as by $G^{\alpha}(V^{\alpha},E^{\alpha})$, $G^{\beta}(V^{\beta},E^{\beta}),\dots$.
These multiple SMNs can be seen as a pair $\mathcal{M}=(g,c)$, where $g={\{ G^\alpha | \alpha \in \{ 1,\dots,m \} \}} $ is a family of graphs denoting the different SMNs and
\begin{equation}
 c={\{E^{\alpha \beta}\subseteq V^\alpha \times V^\beta | \alpha,\beta\in\{1,\dots,m\},\alpha\neq\beta\}}
\end{equation}
is the set of interconnections between the nodes of $G^\alpha$ and $G^\beta$, where $\alpha\neq \beta$. Each element in $g$ is referred to as a layer in $\mathcal{M}$. The elements of $E^\alpha$ are referred to as intralayer links, and the elements of $E^{\alpha \beta}$ are called interlayer links. The interlayer links are also called interlayer node pairs, and the nodes belonging to these pairs are called interlayer nodes.

\textbf{Definition 2: A priori interlayer link.} Given a multiplex network $\mathcal{M}$, if an interlayer link is provided in advance, we refer to it as a priori interlayer link or a priori interlayer node pair. For example, in Fig.~\ref{pic:motivation} (b) or (f), $e_{11}^{\alpha\beta}$, $e_{22}^{\alpha\beta}$, $e_{33}^{\alpha\beta}$, and $e_{44}^{\alpha\beta}$ can be considered as a priori interlayer links, and $(v_1^\alpha,v_1^\beta)$, $(v_2^\alpha,v_2^\beta)$, $(v_3^\alpha,v_3^\beta)$, and $(v_4^\alpha,v_4^\beta)$ can be considered as a priori interlayer node pairs. Meanwhile, the nodes belonging to a priori interlayer node pairs are called matched nodes, and the other nodes are called unmatched nodes. A node pair consisting of two unmatched nodes across different layers can be called an unmatched node pair.

\textbf{Definition 3: unobserved interlayer link.} Given a multiplex network $\mathcal{M}$, if an interlayer link is not provided in advance, we called it unobserved interlayer link. For example, the interlayer links $e_{33}^{\alpha\beta}$ and $e_{44}^{\alpha\beta}$ in Fig.~\ref{pic:motivation} (b) or (f) are the unobserved interlayer links.

\textbf{Definition 4: Common matched neighbor (CMN). }
Given an a priori interlayer node pair ($v^\alpha_i$,$v^\beta_j$), a node $v^\alpha_a$ in layer $\alpha$, and a node $v^\beta_b$ in layer $\beta$, if an intralayer link exists between $v^\alpha_a$ and $v^\alpha_i$ and another intralayer link exists between $v^\beta_b$ and $v^\beta_j$, we can say that the a priori interlayer node pair ($v^\alpha_i$,$v^\beta_j$) is the CMN of nodes $v^\alpha_a$ and $v^\beta_b$. For example, the a priori interlayer node pair ($v^\alpha_5$,$v^\beta_5$) in Fig.~\ref{pic:motivation} (b) or (f) is the CMN of nodes $v^\alpha_5$ and $v^\beta_5$.

\textbf{Definition 5: Network embedding model. }
Given a layer $G^{\alpha}(V^{\alpha},E^{\alpha})$ of the multiplex network $\mathcal{M}$, a network embedding model learns to represent each node $v^\alpha_i \in V^\alpha$ as a low-dimensional vector $\bm{\mathrm{v}}^\alpha_i \in \mathds{R}^d$, where $d$ represents the dimensionality of the latent representation space. For example, the node $v^\alpha_1$ in Fig.~\ref{pic:motivation} (d) is embedded as a two dimensional vector $\bm{\mathrm{v}}^\alpha_1$ by the network embedding model.

\textbf{Definition 6: Mapping function. }
In the method proposed in this paper, each layer is embedded into a single latent representation space. Given a set of interlayer links, the function $\phi$ is defined as a mapping from layer $\alpha$ to layer $\beta$ such that for each interlayer node pair $(v^\alpha_i,v^\beta_j)$, we have $\phi(\bm{\mathrm{v}}^\alpha_i)=\bm{\mathrm{v}}^\beta_j$. In Fig.~\ref{pic:motivation} (d), vectors $\phi(\bm{\mathrm{v}}^\alpha_5)$ and $\phi(\bm{\mathrm{v}}^\alpha_6)$ are mapped by the mapping function from vectors $\bm{\mathrm{v}}^\alpha_5$ and $\bm{\mathrm{v}}^\alpha_6$.

Generally, the perfect mapping function is hard to obtain, as each layer's latent space is unknown to the others~\cite{ZhouFan2018} and the sampled contexts may differ when nodes are embedded into the latent spaces. Most embedding-based methods utilize a priori interlayer links to train an approximate mapping function after achieving the latent representations. After obtaining the approximate mapping function, the unmatched nodes can be unified in a common latent space by this function.

\subsection{Problem Statement}
\begin{figure} [t!]
    \centering
    \includegraphics[width=0.45\textwidth]{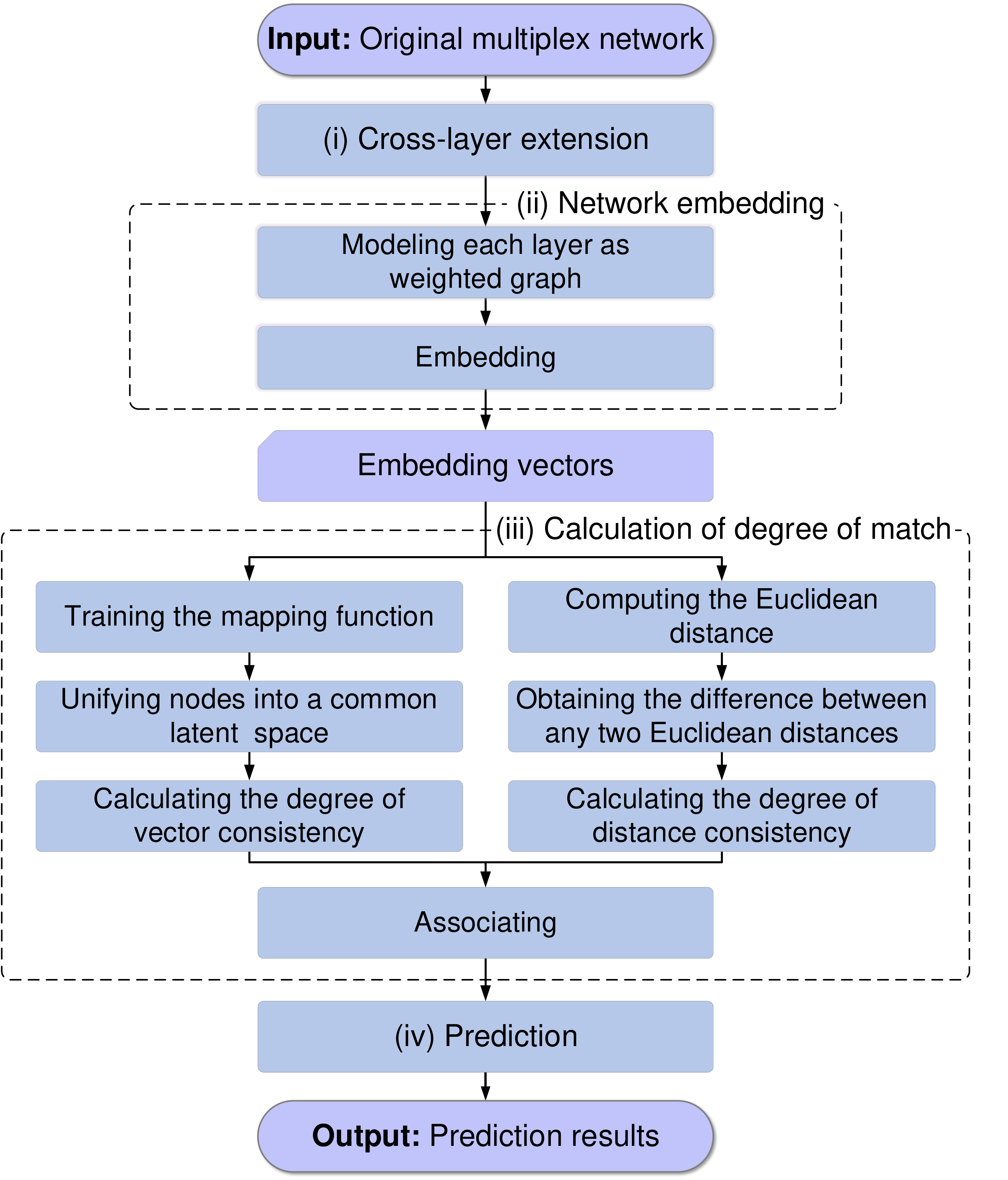}
    \caption{Flowchart of MulCEV. The framework of MulCEV first extends each layer of the multiplex network with the help of a priori interlayer links. Then, it models each layer as a weighted graph and the embedded nodes in different layers into different latent representation spaces to obtain their embedding vectors. Thereafter, MulCEV obtains the degree of match by calculating the degree of vector consistency as well as the distance consistency. Finally, the unobserved interlayer links are predicted.}
    \label{pic:framework}
\end{figure}

Supposing that we have a multiplex network $\mathcal{M}$ with a set of a priori interlayer links, the interlayer link prediction problem is to determine whether any unmatched node pair $v^\alpha_i,v^\beta_j$ chosen from $V^\alpha$ and $V^\beta$ have an interlayer link, i.e., whether the accounts represented by the two unmatched nodes belong to the same person.

Given an unmatched node pair $(u^\alpha_i,u^\beta_j)$ across different layers in the multiplex network $\mathcal{M}$, interlayer link prediction learns a binary function $f : V^\alpha\times V^\beta \rightarrow 0,1$ such that
\begin{equation}
f(u^\alpha_i,u^\beta_j) = \left\{
\begin{array}{ll}
1, & \mbox{if $e^{\alpha\beta}_{ij}$ exist} \\
0, & \mbox{otherwise}
\end{array},
\right.
\label{eq:objectivefunction}
\end{equation}
where $f(u^\alpha_i,u^\beta_j) = 1$  means that there is an interlayer link between unmatched nodes $u^\alpha_i$ and $u^\beta_j$.

It is worth noting that some people may register two or more accounts in a given SMN. For simplicity, we assume that these accounts belong to different individuals.

\section{Proposed Framework}
The proposed framework (shown in Fig.~\ref{pic:framework}) is an algorithm consisting of four main steps: (i) cross-layer extension, (ii) network embedding, (iii) calculation of the degree of match, and (iv) prediction. We discuss each one in detail in the following sections.

\subsection{Cross-layer Extension}
Given two interlayer node pairs in the multiplex network, it is usually true that they have an intralayer link in one layer if there exists a connection in the other layer~\cite{bayati2009algorithms}. Many studies about SMNs found this phenomenon. For example, Ref.~\cite{Narayanan2009} shows that the users on Twitter and Flickr have similar friend relationships; Zhou et al.~\cite{ZhouXiaoping2016} analyzed users on two famous Chinese SMNs, RenRen.com and Sina Microblog, and found that a large percentage of users' friends in Sina Microblog concurred in RenRen.com. Similar foundations have been discovered in Refs.~\cite{JainParidhi2013,zhang2015multiple,zhang2015community,KongXiangnan2013}.
The cross-layer extension is to leverage a priori interlayer links to extend the intralayer links in each layer of the multiplex network, as shown in Fig. 9 in Appendix C.

Given a multiplex network $\mathcal{M}$ with two layers $G^\alpha$ and $G^\beta$, a priori interlayer link set $\Phi$, and intralayer link sets $E^\alpha$ and $E^\beta$, the extended network $\widetilde{G}^\alpha$ of layer $\alpha$ can be described as
\begin{equation}
\begin{array}{ll}
\widetilde{E}^\alpha= & E^\alpha \cup \{(v^\alpha_i,v^\alpha_j)|(v^\alpha_i,v^\beta_a)\in\Phi,\\
&(v^\alpha_j,v^\beta_b)\in\Phi,(v^\beta_a,v^\beta_b)\in E^\beta\},
\end{array}
\label{eq_cro_extension}
\end{equation}
referring to Ref.~\cite{ManTong2016-IJCAI}. The extended network $\widetilde{G}^\beta$ of layer $\beta$ is similar to the above equation. Note that it is not essential to perform cross-network extension.

\subsection{Network Embedding}
In this step, we first provide a method to compute the relationship strengths between nodes with intralayer links such that each layer can be modeled as a weighted graph; then, the nodes in different layers are embedded as low-dimensional vectors in separate latent representation spaces, as shown in Fig. 10 in Appendix C.

During network embedding, nodes that are ``close'' to each other in the network are embedded in such a way that they have similar vector representations~\cite{cai2018comprehensive}. This property is used for unobserved interlayer node pair searching in the unified latent space in Refs.~\cite{ManTong2016-IJCAI,LiuLi2016,ZhouFan2018,ZhouXiaoping2018-IEEE,ren2019meta}. Better embedding may result in improved interlayer link prediction performance. How is it determined whether two nodes are ``close''? Various scholars have proposed different methods. DeepWalk~\cite{PerozziBryan2014} leverages a truncated random walk to generate a set of node sequences for learning the representations. This method considers nodes with intralayer links to be close. LINE~\cite{TangJian2015} uses the notion of first- and second-order proximities to measure closeness, where first-order proximity refers to intralayer links and second-order proximity refers to two nodes having common neighbors. M-NMF~\cite{wang2017community} further incorporates the community structure into network embedding to improve the similarity of node vectors within a community. It is not difficult for two unmatched nodes in different layers to have CMNs within the same community. However, if unmatched nodes have CMNs from different communities, it is more likely that these represent the unobserved interlayer node pairs~\cite{liu2019structural}. Preserving the community structure in the interlayer link prediction task slightly differs from the single network analysis task.
These embedding methods often model different layers as unweighted graphs. However, different intralayer node pairs may have different relationship strengths. For example, if a boy has only one friend, the friendship between him and his companion is highly likely to be closer than that between those who have many friends. To distinguish between relationship strengths, the intralayer connection between nodes should be weighted. We use a weighted-embedding method that embeds each layer of the multiplex network in the form of weighted graphs based purely on the network structure.

The strength of the relationship between two nodes in the same layer can usually be characterized by the number of neighbors they have in common. However, the number of intralayer links a node has also affects the strength of its relationship with other nodes, and the degrees of their common neighbors may also affect relationship strength. Considering these three factors, we propose the following formula:
\begin{equation}
w_{ij}=((\sum_{z \in \Gamma(v_i) \cap \Gamma(v_j)} \frac{1}{\log k_z}) \cdot \frac{|\Gamma(v_i) \cap \Gamma(v_j)|}{|\Gamma(v_i) \cup \Gamma(v_j)|}+1)\cdot e_{ij},
\label{eq:edge_weight}
\end{equation}
where $k_z$ is the degree of node $z$, and $\Gamma(\cdot)$ represents the neighbor set of node inside it.

Using the above formula has the following three advantages:
(i) The greater the number of common neighbors between two nodes, the greater their weight;
(ii) when two pairs of nodes have the same number of common neighbors, the node pair with fewer intralayer links will have the higher weight; and
(iii) the smaller the degree of the common neighbors between two nodes, the greater their weight.

After obtaining the weights of all intralayer links, we reference a famous network embedding model named LINE~\cite{ManTong2016-IJCAI}, which is good at preserving both the local and global network structures and is suitable for weighted networks, to update the node representation. For any intralayer link $e^\alpha_{ij}=(v^\alpha_{i},v^\alpha_j)$ in a given layer $\alpha$, the joint probability between node $v^\alpha_i$ and $v^\alpha_j$ is
\begin{equation}
z(v^\alpha_{i},v^\alpha_j)=\frac{1}{1+\mathrm{exp}(-{(\bm{\mathrm{v}}^\alpha_i)}^{\mathrm{T}} \cdot \bm{\mathrm{v}}^\alpha_j)},
\end{equation}
where $\bm{\mathrm{v}}^\alpha_i $ and $\bm{\mathrm{v}}^\alpha_j$ are the low-dimensional vectors of nodes $v^\alpha_{i}$ and $v^\alpha_j$, respectively, which are defined in $\mathds{R}^d$; $z(\cdot , \cdot)$ is a distribution over the space $V^\alpha\times V^\alpha$, and $(\cdot)^{\mathrm{T}}$ is the transposition function. The empirical counterpart of $z(\cdot , \cdot)$ can be defined as $\widehat{z}(\cdot , \cdot)=w^\alpha_{ij}/ W$, where $w^\alpha_{ij}$ is the weight of the intralayer link $e^\alpha_{ij}$ as calculated by Eq.~(\ref{eq:edge_weight}), and $W$ is the summation of the weights of all intralayer links. By minimizing the KL-divergence~\cite{manning1999foundations} of $z(\cdot , \cdot)$ and its empirical counterpart $\widehat{z}(\cdot , \cdot)$ over all the intralayer links in the $\alpha$ layer, the LINE model can be inferred. The objective function for embedding is
\begin{equation}
O=\sum_{\forall(v^\alpha_{i},v^\alpha_j)\in E^\alpha} \mathrm{KL}(\widehat{z}(v^\alpha_{i},v^\alpha_j),z(v^\alpha_{i},v^\alpha_j)),
\label{eq:embedding_obj1}
\end{equation}
where the KL-divergence $KL(\cdot,\cdot)$ is a method of measuring the similarity of two distributions. By omitting some constants, the objective function for embedding can be rewritten as
\begin{equation}
O=-\sum_{\forall(v^\alpha_{i},v^\alpha_j)\in E^\alpha} w^\alpha_{ij} ~\mathrm{log}(z(v^\alpha_{i},v^\alpha_j)).
\label{eq:embedding_obj2}
\end{equation}
By minimizing Eq.~(\ref{eq:embedding_obj2}) over all the intralayer links independently, each of the nodes in the given layer $\alpha$ can be represented as a $d$-dimensional vector in the latent representation space with the stochastic gradient descent algorithm. The layer $\beta$ of the multiplex network can be embedded by following the same steps.

\subsection{Calculation of Degree of Match}
For any two unmatched nodes across different layers, we calculate a score according to the MulCEV framework to estimate whether an interlayer link exists between them. We call this score the degree of match, which consists of two parts: the degree of vector consistency, and the degree of distance consistency, as shown in Fig. 11 in Appendix C. The details are as follows.

\subsubsection{Degree of vector consistency}
We leverage the feed-forward multi-layer perceptrons (MLP)~\cite{suter1990multilayer} to learn the mapping function from one layer to another based on the a priori interlayer links. The structure of the MLP used in MulCEV is shown in Fig. 11 (a) in Appendix C. Given each of the a priori interlayer node pairs $(v^\alpha_i,v^\beta_j) \in E^{\alpha \beta}$ and their corresponding embedding vectors $(\bm{\mathrm{v}}^\alpha_i,\bm{\mathrm{v}}^\beta_j)$, we use $\bm{\mathrm{v}}^\alpha_i$ as the input and $\bm{\mathrm{v}}^\beta_j$ as the target output to train the mapping function $\phi$.
The loss function of the MLP is
\begin{equation}
 l(\bm{\mathrm{v}}^\alpha_i,\bm{\mathrm{v}}^\beta_j)=1-\mathrm{cos}(\phi(\bm{\mathrm{v}}^\alpha_i),\bm{\mathrm{v}}^\beta_j),
\label{eq:loss_func}
\end{equation}
where $\mathrm{cos}(\cdot,\cdot)$ is the cosine similarity, and $\phi(\bm{\mathrm{v}}^\alpha_i)$ is the actual output mapped by the MLP. The value of the loss function ranges from 0 to 2. Suppose that we have $n$ a priori interlayer links; then for all a priori interlayer nodes, we use $\bm{\mathrm{A}}^\alpha \in \mathds{R}^{(d,n)}$ and $\bm{\mathrm{A}}^\beta \in \mathds{R}^{(d,n)}$ to represent their respective embedding vector matrices. The goal of training the MLP is to minimize the following cost function:
\begin{equation}
L(\bm{\mathrm{A}}^\alpha,\bm{\mathrm{A}}^\beta)=1-\mathrm{cos}(\phi(\bm{\mathrm{A}}^\alpha),\bm{\mathrm{A}}^\beta;\Theta),
\label{eq:cost_func}
\end{equation}
where $\Theta$ is the collection of all parameters in the mapping function $\phi$.

To obtain the degree of vector consistency, for any given unmatched node pair $(u^\alpha_a,u^\beta_b)$ with their embedding vectors $\bm{\mathrm{u}}^\alpha_a$ and $\bm{\mathrm{u}}^\beta_b$, we map node $u^\alpha_a$ into the latent representation space of the $\beta$ layer according to the mapping function $\phi(\bm{\mathrm{u}}^\alpha_a)$. We then use cosine similarity to compute the degree of vector consistency between $\phi(\bm{\mathrm{u}}^\alpha_a)$ and $\bm{\mathrm{u}}^\beta_b$. The formula can be represented as
\begin{equation}
p(u^\alpha_a,u^\beta_b)=\frac{\phi(\bm{\mathrm{u}}^\alpha_a)^{\mathrm{T}} \cdot \bm{\mathrm{u}}^\beta_b}{||\phi(\bm{\mathrm{u}}^\alpha_a)||~\cdot ~||\bm{\mathrm{u}}^\beta_b||},
\label{eq:matchdegree_gm}
\end{equation}
where $||\cdot||$ represents the 2-norm of the vector within.

\subsubsection{Degree of distance consistency}

As shown in Fig.~\ref{pic:motivation} (d), if we consider only the degree of vector consistency, it may be difficult to obtain good prediction results in some cases, such as the incorrect match of $(v^\alpha_5,v^\beta_6)$. The reason is that a perfect mapping function is difficult to obtain ~\cite{ZhouFan2018}. If we consider the positional relationships between the unmatched nodes and their matched neighbors in the embedding spaces of different layers, we might uncover additional clues for predicting the unobserved interlayer links. We propose a ``distance consistency'' index to measure this relationship, defined as
\begin{equation}
q(u^\alpha_a,u^\beta_b)=\sum_{\substack{\forall(v^\alpha_i,v^\beta_j)\in \Phi, \\ v^\alpha_i \in\Gamma(u^\alpha_a),\\v^\beta_j\in\Gamma(u^\beta_b)}} \mathrm{exp}(-(s^\alpha_{ai} \cdot |s^\alpha_{ai}-s^\beta_{bj}| \cdot s^\beta_{bj})).
\label{eq:matchedegree_dc}
\end{equation}
In Eq.~(\ref{eq:matchedegree_dc}), $\Phi$ represents the set of a priori interlayer links, and $s^\alpha_{ai}$ is the Euclidean distance between unmatched node $u^\alpha_a$ and matched node $v^\alpha_i$. The constraints in the equation indicate that the interlayer node pair $(v^\alpha_i$, $v^\beta_j)$ is the CMN of unmatched nodes $u^\alpha_a$ and $u^\beta_b$. Suppose that matched node pair $(v^\alpha_i,v^\beta_j)$ is the CMN of the unmatched nodes $u^\alpha_a$ and $u^\beta_b$; then $|s^\alpha_{ai}-s^\beta_{bj}|$ can measure the degree of similarity between $s^\alpha_{ai}$ and $s^\beta_{bj}$. If the value of $|s^\alpha_{ai}-s^\beta_{bj}|$ is close to 0, the Euclidean distances $s^\alpha_{ai}$ and $s^\beta_{bj}$ are deemed to be consistent; otherwise, $s^\alpha_{ai}$ and $s^\beta_{bj}$ are deemed to be inconsistent. Using $s^\alpha_{ai}$ and $s^\beta_{bj}$ to multiply $|s^\alpha_{ai}-s^\beta_{bj}|$ distinguishes the influence of the CMNs on the degree of distance consistency. The closer the Euclidean distance of an unmatched node and its CMN, the greater the influence of this CMN. $(s^\alpha_{ai} \cdot |s^\alpha_{ai}-s^\beta_{bj}| \cdot s^\beta_{bj})$ can be transformed by the sigmoid function to ensure that its value is between 0 and 1. In addition, for an unmatched node pair, the larger the value of $|s^\alpha_{ai}-s^\beta_{bj}|$, the smaller the distance consistency should be, which is reflected by the exponential function $\mathrm{exp}(- (\cdot))$ in the formula.

In summary, the degree of distance consistency has the following characteristics:

(i) the greater the number of CMNs, the greater the degree of distance consistency;

(ii) the smaller the Euclidean distance between an unmatched node and its CMN, the greater the influence of this CMN on the degree of distance consistency;

(iii) the smaller the difference between two Euclidean distances formed by unmatched nodes across different layers and their CMNs, the larger the degree of distance consistency.

\subsection{Prediction}

After obtaining the degrees of vector consistency and distance consistency for unmatched node pair $(u^\alpha_a,u^\beta_b)$, we associate these two types of consistency to calculate the final degree of match, as shown in Fig. 12 in Appendix C. The formula of associating the two types of consistency can be represented as
\begin{equation}
r(u^\alpha_a,u^\beta_b)=\delta \cdot p(u^\alpha_a,u^\beta_b)+(1-\delta) \cdot q(u^\alpha_a,u^\beta_b),
\label{eq:matchedegree_pair}
\end{equation}
where $\delta$ is a control parameter that takes a value from 0 to 1. If $\delta=0$, the degree of match is only related to the distance consistency, whereas if $\delta=1$, the degree of match is only related to the vector consistency.

For any node $u^\alpha_a$ in layer $\alpha$, we can calculate its degree of match with all unmatched nodes in layer $\beta$. We can then predict an interlayer link by identifying the counterpart node in layer $\beta$ that has the highest degree of match with node $u^\alpha_a$ or offer a top-$N$ list of nodes in layer $\beta$ as potential counterparts of node $u^\alpha_a$.

\subsection{Optimization and Time Complexity}
To reduce the time complexity, we optimized the calculation of the degrees of vector consistency and distance consistency. Through optimization, the degree of match for all unmatched node pairs can be obtained by
\begin{equation}
\bm {\mathrm{R}}=\delta \cdot \bm {\mathrm{P}}+(1-\delta) \cdot \bm {\mathrm{Q}},
\end{equation}
where $\bm {\mathrm{P}}$ and $\bm {\mathrm{Q}}$ are the matrix for the degree of vector and distance consistency, respectively. The specific calculations $\bm {\mathrm{P}}$ and $\bm {\mathrm{Q}}$ are detailed in Appendix D.

The time complexity of the four steps are $O(n^2)$, $O(O(|\Phi|(d\iota+\langle k\rangle^2))$, $O(kdn+n_u^3d/\varsigma+nn_u^2d/\varsigma)$, and $O(Nn_u^2)$, where $\varsigma$ is the number of computational nodes~\cite{lee1997IO} and $N$ is the size of top-$N$ list. The details of time complexity analysis can be found in Appendix E.

\section{Experiments}
In this section, we first describe the experiment configurations and then compare the proposed framework with baseline methods using three synthetic and four real-world datasets.

\subsection{Experimental Configurations}
We used three synthetic and four real-world multiplex network datasets in our experiments. The synthetic networks are Erd\H{o}s-R\'{e}nyi~\cite{erdHos1960evolution} (ER) random networks, Watts-Strogatz~\cite{watts1998collective} (WS) small-world networks, and Barab\'{a}si-Albert~\cite{Barabasi1999-BA} (BA) networks. The four real-world datasets are Foursquare--Twitter (FT)~\cite{ZhangJiawei2015-IJCAI}, DBLP\_DataMining-DBLP\_MachineLearning (DBLP)~\cite{tang2008extraction,liu2019structural}, Higgs\_Friendships-Higgs\_Mention (Higgs-FSMT), and Higgs\_Friendships-Higgs\_Retweet (Higgs-FSRT)~\cite{de2013anatomy}. Meanwhile, we used DeepLink~\cite{ZhouFan2018}, IONE~\cite{LiuLi2016}, ONE~\cite{LiuLi2016}, IONE-D~\cite{liu2019structural}, BootEA~\cite{sun2018bootstrapping}, PALE~\cite{ManTong2016-IJCAI}, MAH~\cite{TanShulong2014-AAAI}, MAG~\cite{TanShulong2014-AAAI}, and CRW~\cite{ZhangJiawei2015-IJCAI} as baselines. In addition, we employed $Precision@N$ ($P@N$)~\cite{LiuLi2016,ShuKai2017}, F-measure ($F1$)~\cite{ShuKai2017}, and $MAP$~\cite{ShuKai2017} as the metrics to evaluate the performance of all methods.
The details of each dataset, baselines, and the other experimental configurations can be found in Appendix F.

\subsection{Experimental Results}
In this subsection, we investigate the effects of the control parameter $\delta$ and compare the baseline methods with the proposed method.

\subsubsection{Effect of control parameter $\delta$}
In Eq.~(\ref{eq:matchedegree_pair}), the parameter $\delta$ is leveraged to control the proportions of the vector consistency and distance consistency in the final degree of match. We studied the initialization strategy for $\delta$ and its effect on the predicted results through experiments. The results and phenomena are shown in Appendix G. In these experiments, the prediction results of most datasets show best performance with $\delta=0.5$, this value is recommended for new datasets. For more precise values, investigators may consider performing 10-fold cross validation using the a priori interlayer link set, which is divided into training and validation sets.

\subsubsection{Performance with different $@N$ settings}
We evaluate the performance of the baseline methods and the proposed method at different $@N$ settings on real-world datasets FT and DBLP. In the proposed framework, the first step is cross-layer extension. This step, however, is not mandatory. We call the version without cross-network extension MulCEV, and the version with cross-network extension MulCEV-Ex.

\begin{figure} [!t]
    \centering
    \includegraphics[width=0.5\textwidth]{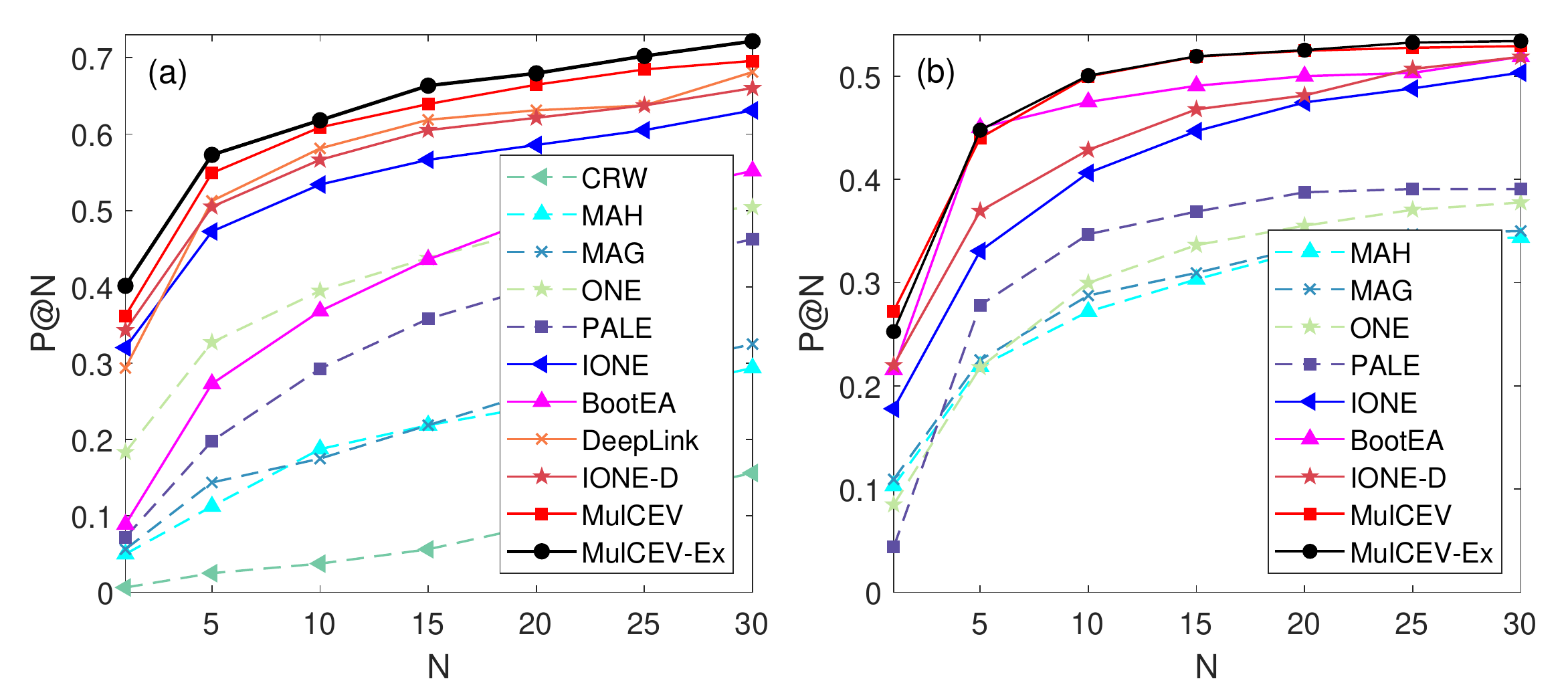}
    \caption{Comparison between baselines and our proposed methods for different $@N$ settings on datasets FT and DBLP.}
    \label{pic:result_diffpatn}
\end{figure}

Referring to Refs.~\cite{liu2019structural} and~\cite{ZhouFan2018}, we set $90.0\%$ of the interlayer links as the training set and the rest as the test set. Figure~\ref{pic:result_diffpatn} displays the precision of the baseline methods and the proposed method under this setting. From the figure, we can see that MulCEV-Ex achieved the highest precision for all $@N$ settings. On the FT dataset, the precision increased by a maximum of $10.8\%$ and an average of $5.8\%$ over DeepLink, the best of the baseline methods. On the DBLP dataset, the precision increased by a maximum of $3.7\%$ and an average of $2.3\%$ over BootEA. MulCEV achieved the second-highest performance, for a maximum increase of $6.8\%$ and $5.7\%$ on the two respective datasets compared with the best of the baseline methods. In contrast with other methods based on network embedding, our method further considers distance consistency with CMNs. The results imply that the distance consistency provides more clues and better facilitates the prediction of interlayer links. MulCEV-Ex was better than MulCEV under most settings because MulCEV-Ex leverages a priori interlayer links to extend each layer of the multiplex network. The extended layer has more edges to guide the embedding than the non-extended network so that the node positions in the embedding space can better reflect the relationships between nodes. Such advantages are highlighted in the subsequent matching process. The improvement of MulCEV-Ex over MulCEV was greater on the FT dataset than on the DBLP dataset, as the percentage of interlayer nodes is greater in the FT dataset than in the DBLP dataset. The greater the number of interlayer links, the greater the number of intralayer links that can be extended.
\begin{table*}[t!]
\centering
\caption{$MAP$ of different methods.}
\setlength{\tabcolsep}{3pt}
\begin{tabular}{cccccccc}
\hline
\multirow{2}*{\textbf{Datasets}}
&\multicolumn{7}{c}{\textbf{Methods}}\\ \cline{2-8}
& MulCEV-Ex& MulCEV &IONE-D& MAG & IONE& PALE& ONE\\
\hline
FT& \textbf{0.4808}& 0.4511& 0.4228& 0.2251& 0.4005 & 0.1420& 0.2563 \\
DBLP& \textbf{0.3651}& 0.3520& 0.2979& 0.3487& 0.2578& 0.1516& 0.1075 \\
\hline
\end{tabular}
\label{tab:result_map}
\end{table*}

CRW was the lowest-precision method, showing that the traditional link-based prediction method is not as accurate as the network embedding approach. MAH and MAG showed better performance than CRW but were a bit worse than the other methods. This may be because MAH needs hypergraph information, which is often difficult to obtain for an actual SMN and thus must be built using specific methods based on the data obtained, leading to poor performance by MAH. MAG uses a formula for the calculation of node-to-node pairwise weights to build a graph for each SMN and obtains ranking results by the same method as MAH, so its performance is similar to MAH's.

With regard to IONE and its two variants ONE and IONE-D, ONE does not consider the information of the input context of the node, and its performance was not as good as that of IONE. IONE-D, although based on IONE, further considers the impact of community-based structural diversity, and so it exhibited better performance than IONE. PALE does not consider the input and output contexts of the node separately; its performance was not as good as IONE's. It is noteworthy that IONE and PALE are two classical methods based on network embedding; IONE embeds all layers into a common latent space, and PALE embeds each layer into a unique space. On the two datasets, MulCEV increased the precision by an average of $7.0\%$ and $6.8\%$, respectively, over IONE and an average of $28.4\%$ and $15.8\%$, respectively, over PALE.

Compared with other methods, BootEA showed superior performance on the DBLP dataset but worse performance on the FT dataset. This may be because during the process of predicting interlayer links, BootEA iteratively labels potential node pairs as training data to overcome the lack of a sufficiently large training set. The percentage of interlayer nodes in the DBLP dataset is smaller than that in the FT dataset. Through automatic labeling, more training data are provided in the DBLP dataset, thereby allowing the advantages of BootEA to be better reflected. DeepLink gave the best performance of the baselines because it utilizes dual learning for the pretraining of the mapping function.

With all methods, we observe that as $N$ increases, the precision of the various methods also increases. This is because $@N$ denotes the number of potential matches recommended by different methods for each unmatched node. The greater the value of $N$, the higher the number of candidate matches and the higher the probability of success in finding the correct match.

We also investigated the ranking performance of our suggested methods and some baselines with the $90.0\%$ training ratio; Table~\ref{tab:result_map} shows the results. The highest value for each dataset is in boldface. We can see that MulCEV-Ex outperformed all the comparison methods, and MulCEV was better than all the baseline methods. This observation further demonstrates the effectiveness and merits of the proposed framework.
\begin{figure} [t!]
    \centering
    \includegraphics[width=0.5\textwidth]{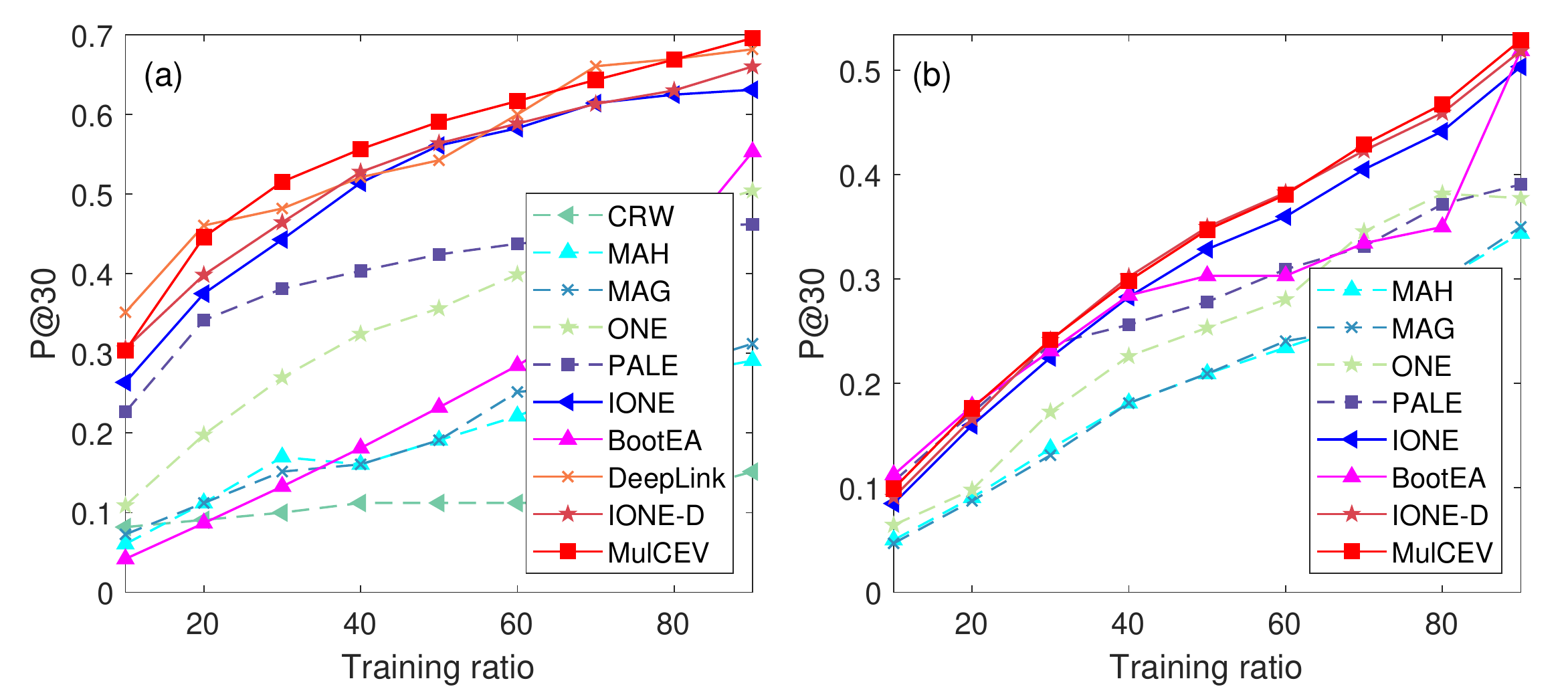}
    \caption{Comparison between baselines and MulCEV for different training ratios on datasets FT and DBLP.}
    \label{pic:result_difftr}
\end{figure}

It is worth noting that of all the methods, MulCEV-Ex is the only one that extends the intralayer links by interlayer links. Therefore, in order to conduct the comparison of the different methods under the same conditions to the extent possible, in the subsequent experiments we excluded MulCEV-Ex and used only MulCEV for the comparisons with the baselines.

\subsubsection{Effect of training ratio}
We evaluated the performance of the baselines and MulCEV under different settings for the training ratio. We set training ratios of $10\%$ to $90\%$ in $10\%$ increments; $N=30$. Figure~\ref{pic:result_difftr} displays the $P@30$ of the baselines and MulCEV under these settings.
\begin{figure} [t!]
    \centering
    \includegraphics[width=0.5\textwidth]{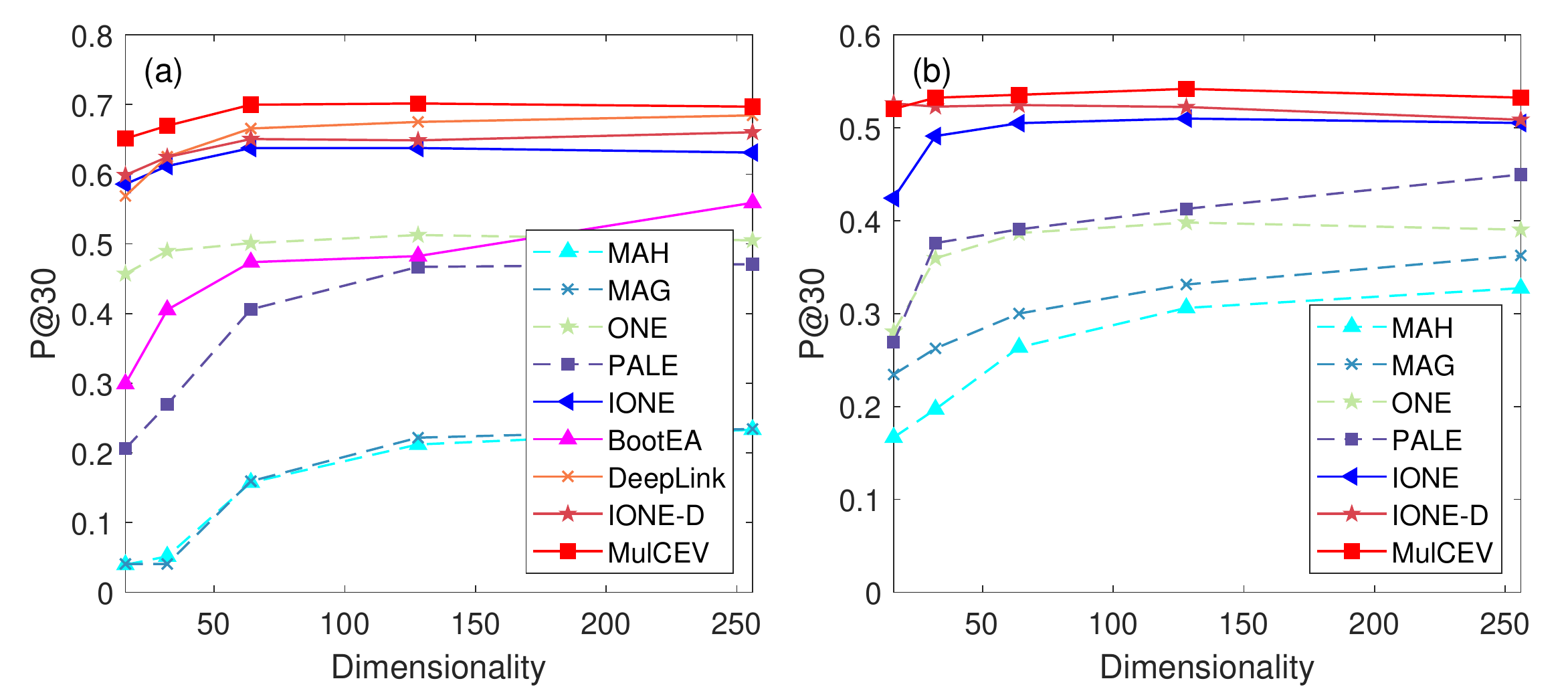}
    \caption{Comparison between baselines and MulCEV for different dimensionalities on datasets FT and DBLP.}
    \label{pic:result_diffd}
\end{figure}

From the figure, we can see that the proportion of interlayer links used for training markedly affected the performance of all of the methods. For each method, $P@30$ increased with the training ratio. This is because the greater the training ratio, the greater the quantity of training data. For the methods that embed each layer into a unique latent space, there are more inputs to learn the mapping function; for the method embedding all layers into a common latent space, there are more inputs to align nodes to the common embedding space. Moreover, the rankings of the performance of all methods on the two datasets are similar to those under various $@N$ settings. The reasons are the same as those illustrated in Fig.~\ref{pic:result_diffpatn}. In particular, MulCEV achieved the highest precision for almost all training ratios. On the FT dataset, $P@30$ increased by a maximum of $4.8\%$ and an average of $0.8\%$ over DeepLink, the best of the baseline methods. On DBLP, it increased by a maximum of $1.0\%$ and an average of $0.4\%$ over the best baseline, IONE-D. These observations demonstrate the effectiveness and merits of the proposed method.

\subsubsection{Effect of embedding dimensionality}
We also evaluated the performance of the network embedding learning-based baseline methods and MulCEV using representations of different dimensionalities $d$. We set $d$ to $16$, $32$, $64$, $128$, and $256$; the training ratio was $90.0\%$, and $N=30$. Figure~\ref{pic:result_diffd} displays the $P@30$ values for the baselines and MulCEV under these settings.

From Fig.~\ref{pic:result_diffd}, we can see that the rankings of the performance of all methods on the two datasets are similar to the rankings under various $@N$ settings. The reasons are the same as those illustrated in Fig.~\ref{pic:result_diffpatn}. In particular, MulCEV achieved the highest precision for almost all dimensionalities. On the FT dataset, $P@30$ increased by a maximum of $8.2\%$ and an average of $4.0\%$ over DeepLink, the best of the baseline methods. On DBLP, it increased by a maximum of $2.3\%$ and an average of $1.2\%$ over the best baseline. These observations demonstrate the effectiveness and merits of the proposed framework. Moreover, we can see that MulCEV, DeepLink, IONE-D, and IONE achieved their best performance on the FT dataset with $d=128$, and on the DBLP dataset with $d=64$. Other methods needed more dimensions to achieve their best performance. It is well known that the computational complexity of learning algorithms is highly dependent on the dimensionality of the embedding space: The lower the dimensionality, the lower the computational complexity. These results again demonstrate the effectiveness and merits of the proposed framework.

\subsubsection{Effect of iteration count}
\begin{figure} [t!]
    \centering
    \includegraphics[width=0.5\textwidth]{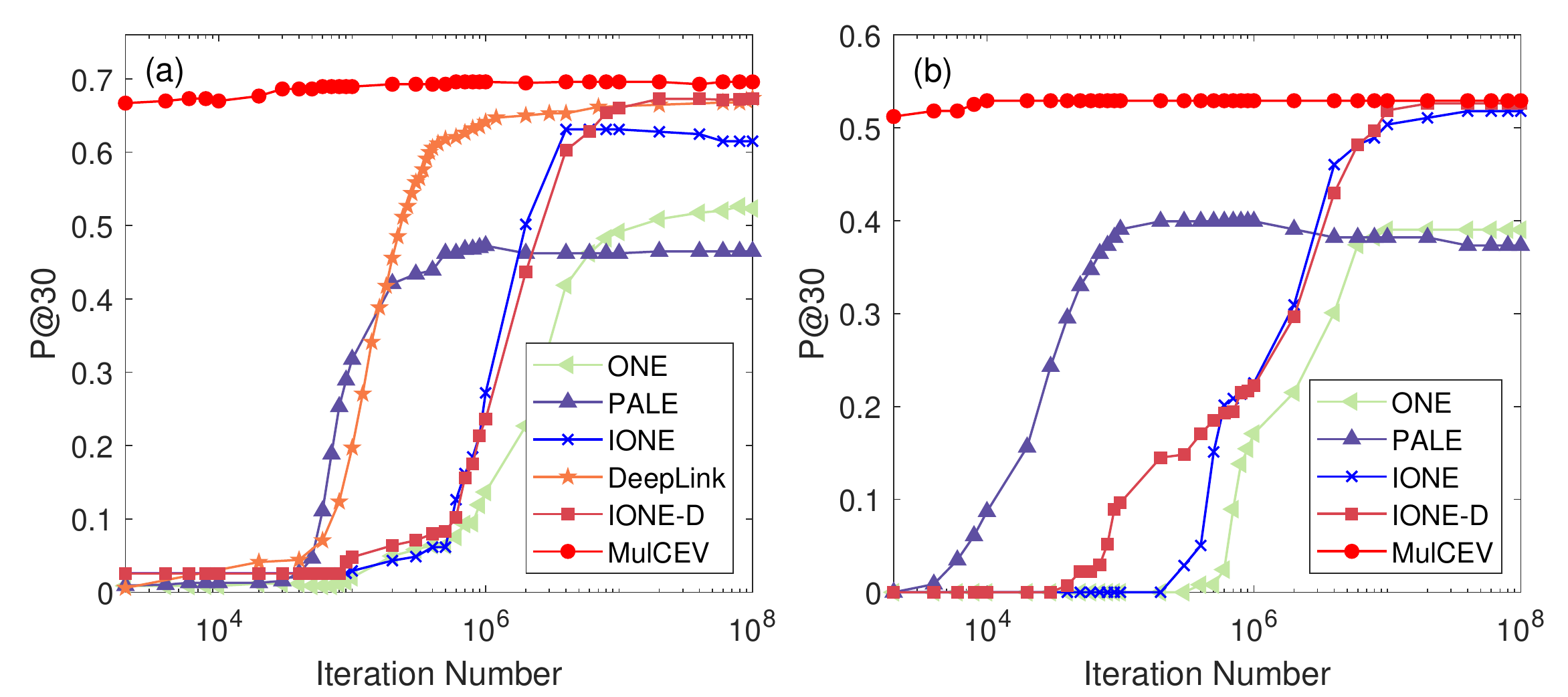}
    \caption{Comparison between baselines and MulCEV for different training iteration counts on datasets FT and DBLP.}
    \label{pic:result_diffitter}
\end{figure}

The number of training iterations needed for a prediction method to converge is another important factor to consider in evaluating these methods. Referring to Refs.~\cite{liu2019structural,ZhouFan2018}, we set the training ratio to $90.0\%$ and set $N=30$ to execute the experiments for the evaluation of the baselines and MulCEV with different numbers of iterations. Figure~\ref{pic:result_diffitter} displays $P@30$ for the baselines and MulCEV under these settings.

From the figure, we can see that the rankings of the performance of all methods on the two datasets are similar to those under various $@N$ settings. The reasons are the same as those illustrated in Fig.~\ref{pic:result_diffpatn}. Meanwhile, MulCEV achieved the highest precision at almost all iteration counts. In particular, it achieved competitive results at very low training iteration counts; at 2000 iterations, it achieved $P@30$ values of $0.667$ and $0.512$ on the two respective datasets.
In contrast, the $P@30$ values of all the baselines were close to zero. This is because the degree of match for MulCEV consists of two parts, degree of vector consistency and degree of distance consistency. The latter is calculated in advance (before the mapping function is learned) and provides some clues for making the predictions. Thereafter, the mapping function improves as the number of iterations increases, further improving the prediction performance.

DeepLink and PALE converged at similar iteration counts, around $10^5$. This is probably because these methods are based on similar concepts. They embed each layer of the multiplex network into a unique latent space and then use MLP to learn the mapping function and complete the matching. The convergent iteration counts for IONE and its two variant methods ONE and IONE-D are also similar, all of them converging between $10^6$ and $10^7$. The reason is the same as that for DeepLink and PALE.
Moreover, we can see that DeepLink and PALE converge to their best performance sooner than IONE and its variant methods. This is probably because IONE and its variants need to learn the context information for the nodes in each layer, and so they require a greater number of learning rounds to converge. $P@30$ for IONE and PALE would decrease at higher iteration counts because they incur overfitting problem.

\subsubsection{Results on more datasets}
\begin{figure} [t!]
    \centering
    \includegraphics[width=0.5\textwidth]{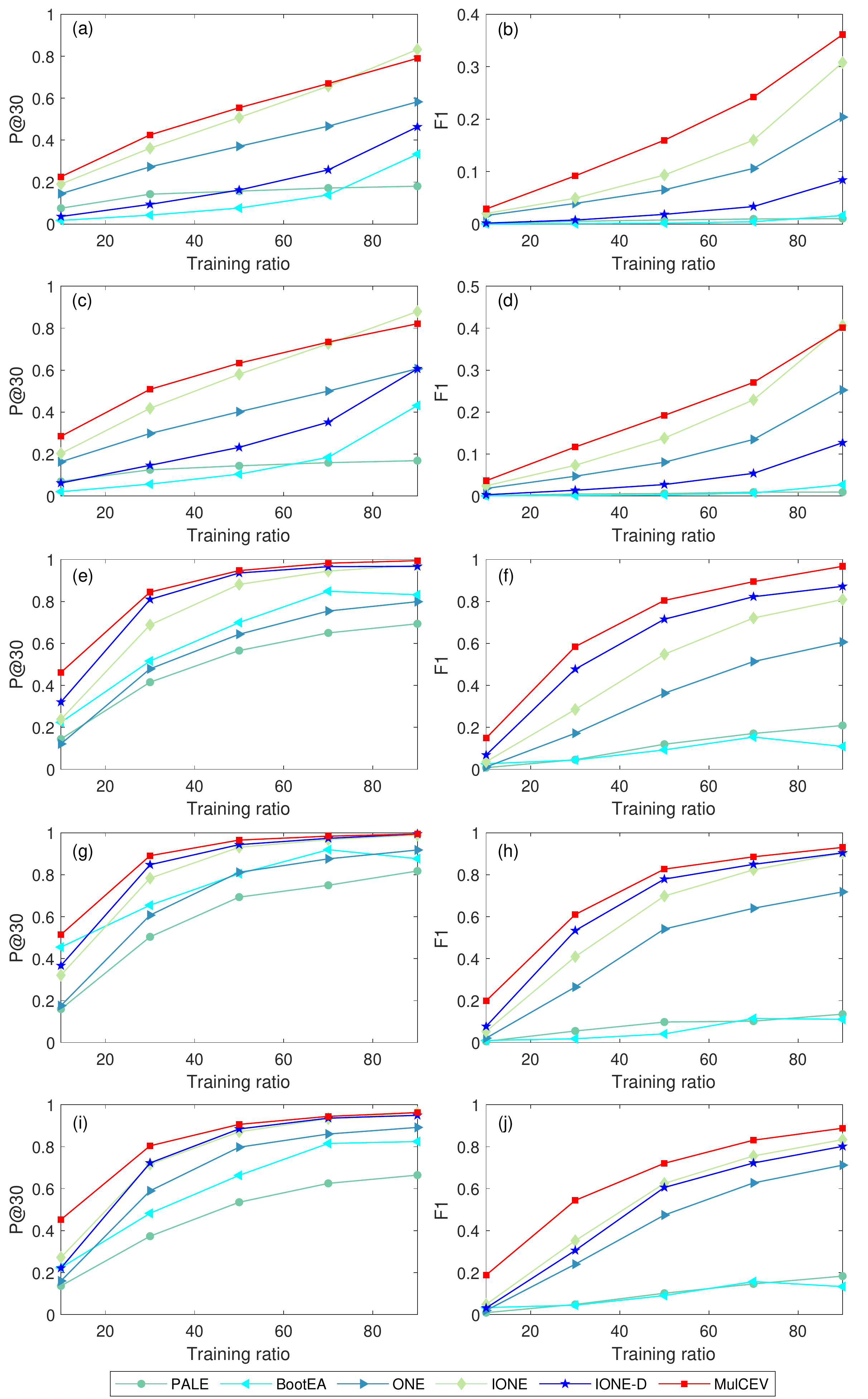}
    \caption{Comparison between baselines and MulCEV on more datasets. The subfigures in the first column are $P@30$ of different training ratios on the datasets of Higgs-FSMT, Higgs-FSRT, ER, WS, and BA.  The subfigures in the second column are $F1$ of different training ratios on these datasets.}
    \label{pic:result_moredatasets}
\end{figure}
To fully evaluate the effectiveness of the proposed approach, we implemented experiments on additional datasets and adopted more metrics. The training ratios in these experiments were varied from $10.0\%$ to $90.0\%$ in increments of $20.0\%$. Fig.~\ref{pic:result_moredatasets} displays the $P@30$ and $F1$ values for the baselines and MulCEV under these settings.
\begin{table*}[htb]
\centering
\caption{Time Cost. }
\label{table}
\setlength{\tabcolsep}{8pt}
\begin{tabular}{cccccc}
\hline
\multirow{2}*{\textbf{Metric}} &\multirow{2}*{\textbf{Datasets} }
& \multicolumn{4}{c}{\textbf{Methods}} \\ \cline{3-6}
&& PALE& BootEA& MulCEV*& MulCEV\\
\hline
\multirow{7}*{$\bm{Time(s)}$}
&FT&434.55& 16765.89&  4669.99&\textbf{216.12}\\
&DBLP&434.93&984.81&3697.31& \textbf{114.60}\\
&Higgs-FSMT&484.81& 15452.64&3106.18&\textbf{296.62}\\
&Higgs-FSRT&404.85& 14112.72&3437.15&\textbf{311.74}\\
&ER&\textbf{4.54}&339.86&48.11&8.49\\
&WS&\textbf{4.07}&274.74&54.50&9.98\\
&BA&\textbf{6.88}&345.45&50.77&8.55\\
\hline
\multicolumn{6}{l}{Note: MulCEV* is the MulCEV before optimization.}\\
\end{tabular}
\label{tab:timecost}
\end{table*}

From Fig.~\ref{pic:result_moredatasets}, we see that MulCEV outperformed the baseline methods in almost all conditions for all data sets, both for $P@30$ and $F1$. This demonstrates the effectiveness and merit of the proposed method. Meanwhile, we observed that $P@30$ and $F1$ of IONE-D were lower than those of IONE on the Higgs-FSMT and Higgs-FSRT datasets, which are different from that of the other datasets in Fig.~\ref{pic:result_difftr} and Fig.~\ref{pic:result_moredatasets}. This may be attributed to the interactions of mention or reply being generated when users discuss a wide range of topics. Such discussions are not necessarily restricted to some communities. IONE-D incorporates the community-based structural diversity into IONE and has the side effect of prediction on the above two datasets. This indirectly demonstrates the robustness of our proposed method, which improves the prediction performance by fully using the information in the latent representation spaces.

\subsubsection{Effect of embedding method}
\begin{figure} [t!]
    \centering
    \includegraphics[width=0.5\textwidth]{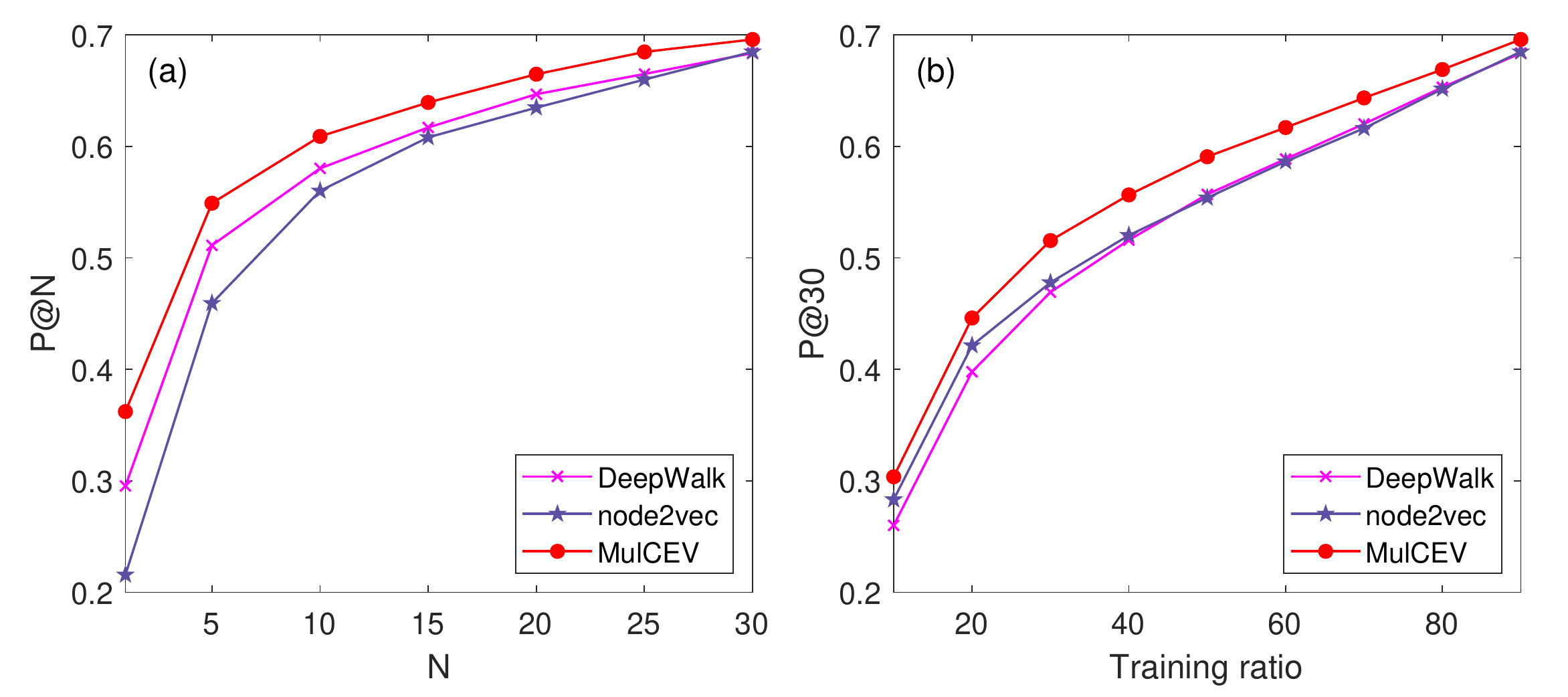}
    \caption{Comparison of different embedding methods. (a) Precision of different $@N$ settings on the dataset FT, (b) $P@30$ of different training ratios on the dataset FT.}
    \label{pic:result_diffemmethods}
\end{figure}
To evaluate the weighted-embedding method used in Section IV, we compared it with two commonly used network embedding methods: DeepWalk~\cite{PerozziBryan2014} and node2vec~\cite{grover2016node2vec}. For DeepWalk, we set the number of walks per node to $20$, the walk length to $80$, and the window size to $5$; for node2vec, we empirically set $q = 0.5$ and $p = 2$.

We compared the weighted-embedding method and the comparison methods under various $@N$ settings and various training ratios on the FT dataset. Figure~\ref{pic:result_diffemmethods} displays the results. As can be seen in the figure, MulCEV achieved the highest precision at almost all $@N$ settings and training ratios. At the different $@N$ settings, $P@N$ increased by a maximum of $6.7\%$ and an average of $2.9\%$ over DeepWalk, the better of the two comparison embedding methods. At the different training ratios, $P@30$ increased by a maximum of $3.8\%$ and an average of $2.7\%$ over node2vec, the better of the two comparison embedding methods. These observations demonstrate the effectiveness and merits of the weighted-embedding method.

\subsubsection{Time cost}

To reduce the time complexity, we adopt matrix multiplication to optimize the process for calculating the different types of consistencies. We compare the time cost of the unoptimized MulCEV, i.e., MulCEV*, and the MulCEV with PALE and BootEA that can be performed on similar hardware and software environments. We executed the experiments on four real-world and three synthetic datasets with a $90.0\%$ training ratio; all experiments were performed on a server with an Intel(R) Xeon(R) Gold 6130 CPU, 256 GB memory, and NVIDIA Tesla V100 GPU, with Python 3.6 and TensorFlow 1.13.1.

Table~\ref{tab:timecost} summarizes the results of these experiments. We see that BootEA required the maximum time, followed by MulCEV*, PALE, and MulCEV on the real-world datasets while BootEA required the maximum time, followed by MulCEV*, MulCEV, and PALE on the synthetic datasets. BootEA employs an iterative training strategy, which updates the predicted oriented embeddings by labeling likely interlayer links and adding them to the a priori interlayer link set iteratively; this method requires an extra $O(n)$ of time over the other algorithms without the iterative training strategy. Although MulCEV* calculates the vector consistency in a manner similar to that of PALE, it consumes more time than PALE because it also computes the distance consistency. However, after optimization, our proposed MulCEV framework effectively reduces the time complexity and requires less time than PALE on the real-world datasets even when both vector consistency and distance consistency are calculated. On the synthetic datasets, MulCEV requires more time than PALE. This is because the number of nodes in each layer of these synthetic multiplex networks is small (about 1000), and both PALE and MulCEV spend a little time. The time saved in calculating vector consistency through optimization is not enough to offset the time spent on calculating the distance consistency.

\section{Conclusion}
We have proposed a framework called MulCEV to predict the interlayer links in a multiplex network. This framework makes full use of the information in the latent representation space through vector consistency and distance consistency. Distance consistency leverages CMNs of the unmatched nodes across different layers as references to provide additional clues for interlayer link prediction. In addition, we modeled the layers as weighted graphs to obtain representation for network embedding so that the higher the strength of the nodes' relationships, the more similar their embedding vectors in the latent representation space. To reduce the time complexity, we adopted matrix multiplication to optimize the process for calculating the degree of match. Experiments on four real-world and three synthetic multiplex network datasets demonstrated that the proposed MulCEV framework markedly outperforms several state-of-the-art methods.

In summary, the proposed framework further improves the accuracy of the network-based embedding method for dealing with interlayer link prediction, especially when the number of training iterations is low. The framework can effectively associate the accounts belonging to the same user across different SMNs solely by leveraging network structure attributes in the absence of attribute information such as username, age, or published content. Such an association can be used to establish patterns of law violations by cybercriminals, improve the understanding of information diffusion across SMNs, and provide support for criminal investigations and evidence collection through SMNs.

Although MulCEV has a better performance, there still exist some limitations. First, it only uses the structural information for network embedding while neglecting the rich attributes of nodes or links. Second, the multiplex network MulCEV focuses on is a fixed one. However, SMNs evolve over time and many application scenarios require predictions to be quickly made for unseen nodes or entirely new subnetworks. In the future, we plan to further explore more reasonable embedding methods to capture the network structure as well as attributes and make predictions in scenarios in which the number of nodes is dynamically increased.

\bibliographystyle{IEEEtran}
\bibliography{References}

\begin{appendices}
\section{More Related Work}
The problem of interlayer link prediction in the multiplex network is typically solved by leveraging feature or structure information accessed from the multiple SMNs~\cite{ShuKai2017}. Early studies focused on feature information; they analyzed user profiles, location trajectories, and user-generated content to link nodes across different SMNs belonging to the same user. Profile features included username, image, position, birthday, job, and experience, among others~\cite{ZhaoDongsheng2018}. The authors of Refs.~\cite{Zafarani2009,Perito2011,Zafarani2013,LiuJing2013-WSDM,li2019matching} explored ways of using usernames for prediction. References~\cite{Carmagnola2009,Iofciu2011,AbelFabian2013,Goga2013,Cortis2013, MuXin-KDD2016} considered various profile attributes to improve prediction performance. With the rapid development of SMNs, many users began using different usernames in different SMNs for security reasons. Meanwhile, the accessible profile information among SMNs became increasingly fragmented, unavailable, and disruptive~\cite{fu2020deep}. These SMN characteristics marginalized the traditional profile-based resolutions. The trajectory-based method has been popular since the emergence of the mobile-phone-based Internet. SMN users who wish to announce their location to their friends on some SMN applications can tap a ``check-in'' button to see a list of nearby places and choose the place that matches their location. References~\cite{Riederer2016,chen2018effective,feng2019dplink} focused on these check-in data and used them to link identities. Such trajectory-based methods, however, often face data sparsity problems, and users usually share different locations on different SMNs. User-generated content can reveal some unique characteristics of an SMN user, such as his or her writing style~\cite{ZhengRong2006,narayanan2012feasibility} or footprint~\cite{Goga2013a}. These methods rely heavily on the availability of excellent natural language processing (NLP) techniques and text preprocessing algorithms because user-generated content often includes spoken words, emotion icons, and abbreviations.

Different from feature information, network's structural information is highly accessible and difficult to counterfeit. In addition, a user's friend circle is highly personalized; i.e., few people share the same friend circle~\cite{ZhouXiaoping2016}. Therefore, network-based methods are an ideal solution for the interlayer link prediction problem and have attracted the interest of an increasing number of researchers in recent years. Network-based methods can be divided into non-embedding-based methods and embedding-based methods according to whether network embedding techniques are used. We will introduce non-embedding-based methods as follows.
\begin{table}[!t]
\renewcommand{\arraystretch}{1.3}
\caption{Symbols and notations}
\label{tab:symbol}
\centering
\begin{tabular}{|p{1cm}||p{6cm}|}
\hline
\textbf{Symbol }& \textbf{Description}\\
\hline
\hline
$\mathcal{M}$ & The multiplex network.\\
\hline
$G$ & A SMN which is one layer of $\mathcal{M}$. \\
\hline
$u,v$  & Nodes in $\mathcal{M}$. \\
\hline
$\bm {\mathrm{u}},\bm {\mathrm{v}} $&Embedding vectors of nodes $u$ and $v$ respectively.\\
\hline
$\alpha,\beta$ & Layer indices of $\mathcal{M}$.\\
\hline
$e^\alpha,e^\beta$  & Intralayer links in $G^\alpha$ and $G^\beta$ respectively. \\
\hline
$\bm{\mathrm{e}}$,$\bm{\mathrm{E}}$ & Intralayer links vector and intralayer link matrix respectively.\\
\hline
$e^{\alpha\beta}$  & Interlayer link. \\
\hline
$i,j,a,b$ & Node indices.\\
\hline
$n^\alpha,n^\beta$ & Number of nodes in $G^\alpha$ and $G^\beta$.\\
\hline
$n,m$ & Number of a priori interlayer links and unobserved interlayer links respectively.\\
\hline
 $\Gamma(v_i)$ & Set of neighbors of node $v_i$. \\
\hline
$k_v$ & Degree of node $v$\\
\hline
$p, \bm{\mathrm{P}}$ & Degree of vector consistency and Degree of vector consistency matrix respectively.\\
\hline
$q, \bm{\mathrm{Q}}$ & Degree of distance consistency and Degree of distance consistency matrix respectively.\\
\hline
$r, \bm{\mathrm{R}}$ & Degree of match and Degree of match matrix respectively.\\
\hline
$d$ & Dimensionality of the latent representation space. \\
\hline
$\phi$ & Mapping function. \\
\hline
$w$ & weight of the intralayer link.\\
\hline
$\delta$ & control parameter. \\
\hline
$\Phi$ & Set of a priori interlayer links.\\
\hline
$\Psi$ & Set of unobserved interlayer links.\\
\hline
\end{tabular}
\end{table}

Given the completeness and connectivity of a network structure, two kinds of structural information can be used to solve the interlayer link prediction problem. The first is local network information, which focuses on the one-hop neighborhood (e.g., follower/followee/friend relationships) of the unmatched nodes~\cite{ShuKai2017}. Narayanan and Shmatikov proposed a re-identification algorithm, which was the first method to use a graph-theoretic model based on the node neighborhood to solve this problem~\cite{Narayanan2009}. Later, Korula et al.~\cite{korula2014efficient} computed a similarity score for an unmatched node pair by counting the number of CMNs and then keeping all the links above a specific threshold. To avoid the possible problem of mismatching low-degree nodes in the early phases, only nodes whose degree is higher than a specified threshold are allowed to be matched. Zhou et al.~\cite{ZhouXiaoping2016} proposed a friend-relationship-based user identification (FRUI) algorithm that counts the number of shared friends to calculate the degree of match for all candidate-matched node pairs and chooses pairs that have the maximum value as the final set of matched pairs. Tang et al.~\cite{tang2020interlayer} further investigated the importance of the scale-free property of real-world SMNs for accomplishing interlayer link prediction and proposed a degree penalty principle to calculate the degree of match of all unmatched node pairs. Ren et al.~\cite{ren2019meta} defined a set of meta-diagrams for feature extraction and used greedy link selection for the interlayer link prediction. In Ref.~\cite{zhang2015multiple}, an algorithm is proposed to resolve the one-to-one constraint in the situation of prediction cross-multiple layers. This algorithm matches different layers of the multiplex network by minimizing the friendship inconsistency and selects the candidate node pairs which can lead to the maximum confidence scores across multiple layers.

The second type of structural information is global network information. Zhu et al.~\cite{ZhuYuanyuan2012} transformed the interlayer link prediction problem into a maximum common subgraph problem and maximized the number of intralayer links to obtain a cross-layer mapping. Zafarani and Liu~\cite{zafarani2015user} also explored a solution utilizing global network information. They calculated the Laplacian matrices for each layer and used a matrix optimization method to perform the prediction. The authors of Ref.~\cite{ZhangYutao2015} considered both local and global consistency to match nodes across more than two layers: local consistency for matching nodes across just two layers, and global consistency for dealing with the cases involving more than two layers. In Refs.~\cite{ZhangSi2016-KDD,ZhangSi2019-www}, the authors studied ways of predicting interlayer links in the absence of a priori interlayer links using the global network information.

\section{Table of Symbols and Notations}
Table~\ref{tab:symbol} displays the main symbols and notations used in this paper. We follow the common symbolic conventions, wherein bold uppercase letters denote matrices, bold lowercase letters denote column vectors, and lowercase letters denote scalars.

\section{Example of Different Steps of MulCEV}
\begin{figure} [t!]
    \centering
    \includegraphics[width=0.49\textwidth]{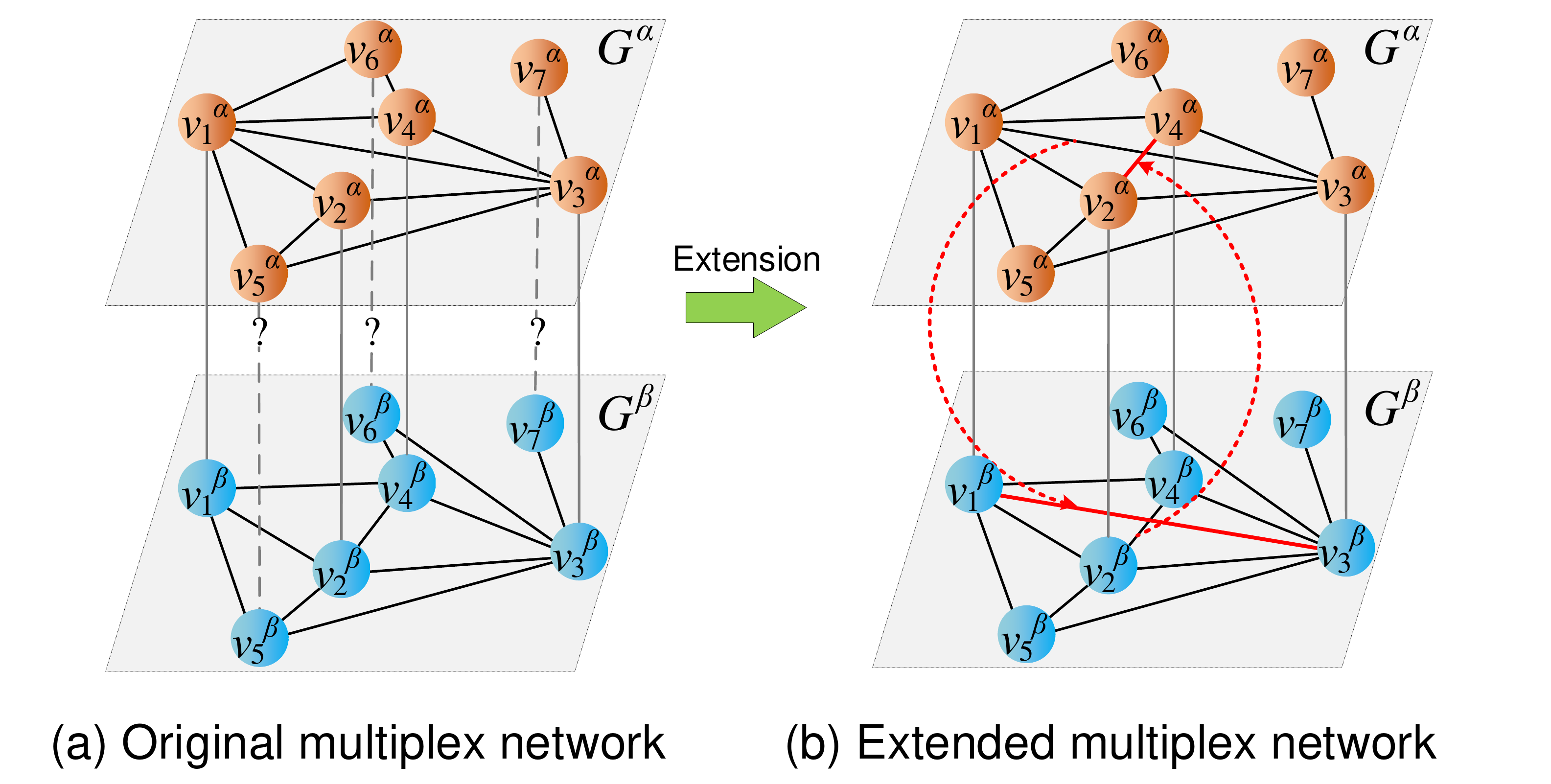}
    \caption{Example of cross-layer extension. (a) Original multiplex network. Given a multiplex network with two layers and each layer having seven nodes, seven interlayer links exist between the nodes across the layers; four of these are a priori interlayer links, whereas the remaining three are unobserved interlayer links. (b) Extended multiplex network. Some of the missing intralayer links between the matched nodes in one layer can be assumed to be present with the assistance of their counterpart nodes in the other layer. For example, there exists an intralayer link $e^\beta_{24}$ between the matched nodes $v^\beta_2$ and $v^\beta_4$; this enables extension of the layer $\alpha$ with the intralayer link $e^\alpha_{24}$ (red line in layer $\alpha$).}
    \label{pic:extension}
\end{figure}

\begin{figure} [t!]
    \centering
    \includegraphics[width=0.49\textwidth]{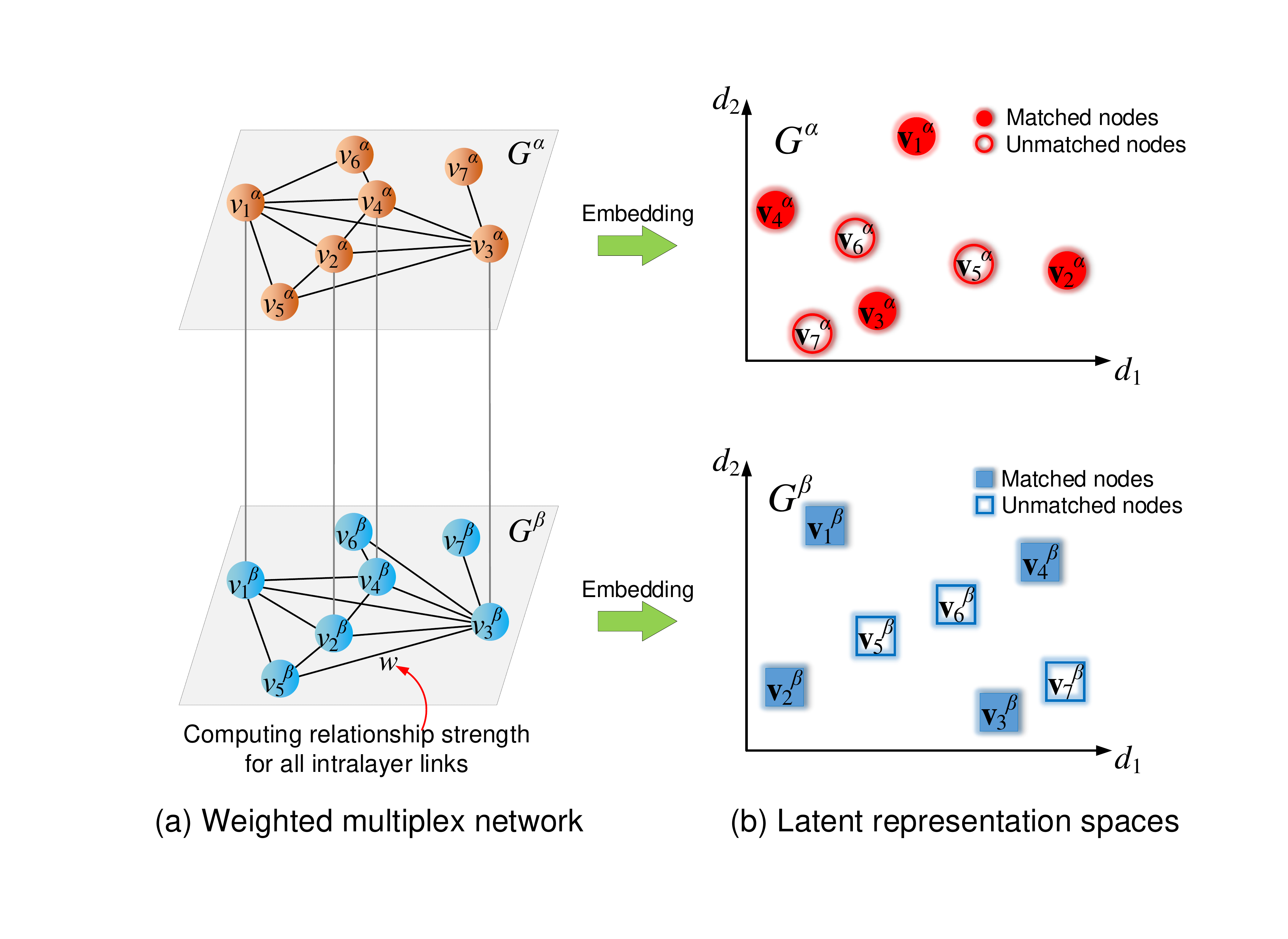}
    \caption{Example of network embedding. (a) Weighted multiplex network: The relationship strengths between the nodes having intralayer links are computed to model each layer as a weighted graph. (b) Latent representation spaces: The nodes in different layers are represented as low-dimensional vectors in separate latent spaces. For example, node $v^\alpha_1$ is represented as a two-dimensional vector $\bm{\mathrm{v}^\alpha_1}$ in the latent space $G^\alpha$.}
    \label{pic:embedding}
\end{figure}

\begin{figure*} [t!]
    \centering
    \includegraphics[width=0.99\textwidth]{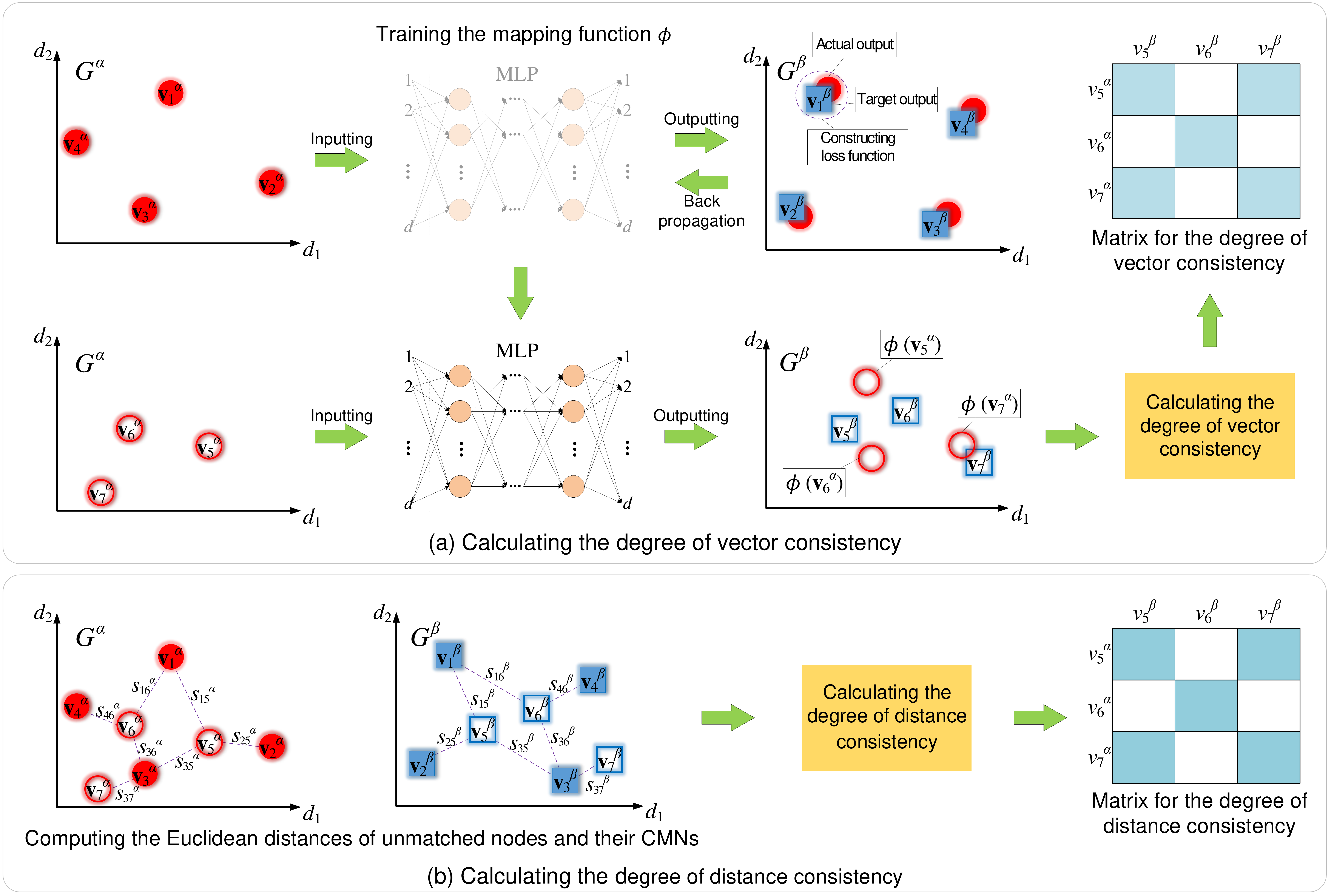}
    \caption{Example of calculating degree of match. (a) Calculating the degree of vector consistency: The embedding vectors of the matched nodes in layer $\alpha$, i.e., $\bm{\mathrm{v}}^\alpha_1$, $\bm{\mathrm{v}}^\alpha_2$, $\bm{\mathrm{v}}^\alpha_3$, and $\bm{\mathrm{v}}^\alpha_4$, are specified as inputs while the embedding vectors of the matched nodes in layer $\beta$, i.e., $\bm{\mathrm{v}}^\beta_1$, $\bm{\mathrm{v}}^\beta_2$, $\bm{\mathrm{v}}^\beta_3$, and $\bm{\mathrm{v}}^\beta_4$, are specified as the target outputs for the MLP to train the mapping function. Then, the embedding vectors of the unmatched nodes in layer $\alpha$, i.e., $\bm{\mathrm{v}}^\alpha_5$, $\bm{\mathrm{v}}^\alpha_6$, and $\bm{\mathrm{v}}^\alpha_7$, are applied as inputs to the trained MLP to obtain their mapped vectors in layer $\beta$, i.e., $\phi(\bm{\mathrm{v}}^\alpha_5)$, $\phi(\bm{\mathrm{v}}^\alpha_6)$, and $\phi(\bm{\mathrm{v}}^\alpha_7$). Thereafter, the cosine similarities between the mapped vectors of the unmatched nodes in layer $\alpha$ and the embedding vectors of the unmatched nodes in layer $\beta$ are calculated to obtain the matrix for the degree of vector consistency. (b) Calculating the degree of distance consistency: The Euclidean distances between the unmatched nodes, i.e., $v^\alpha_5$, $v^\alpha_6$, or $v^\alpha_7$, in layer $\alpha$ and their matched neighbors are computed; similar calculations are performed for the unmatched nodes in layer $\beta$. Then, the differences between pairs of Euclidean distances formed by the unmatched nodes across different layers and their CMNs are calculated. Finally, these differences and the numbers of CMNs between the unmatched nodes across the different layers are combined to obtain the matrix for the degree of distance consistency.}
    \label{pic:degreeofmatch}
\end{figure*}

\begin{figure} [t!]
    \centering
    \includegraphics[width=0.49\textwidth]{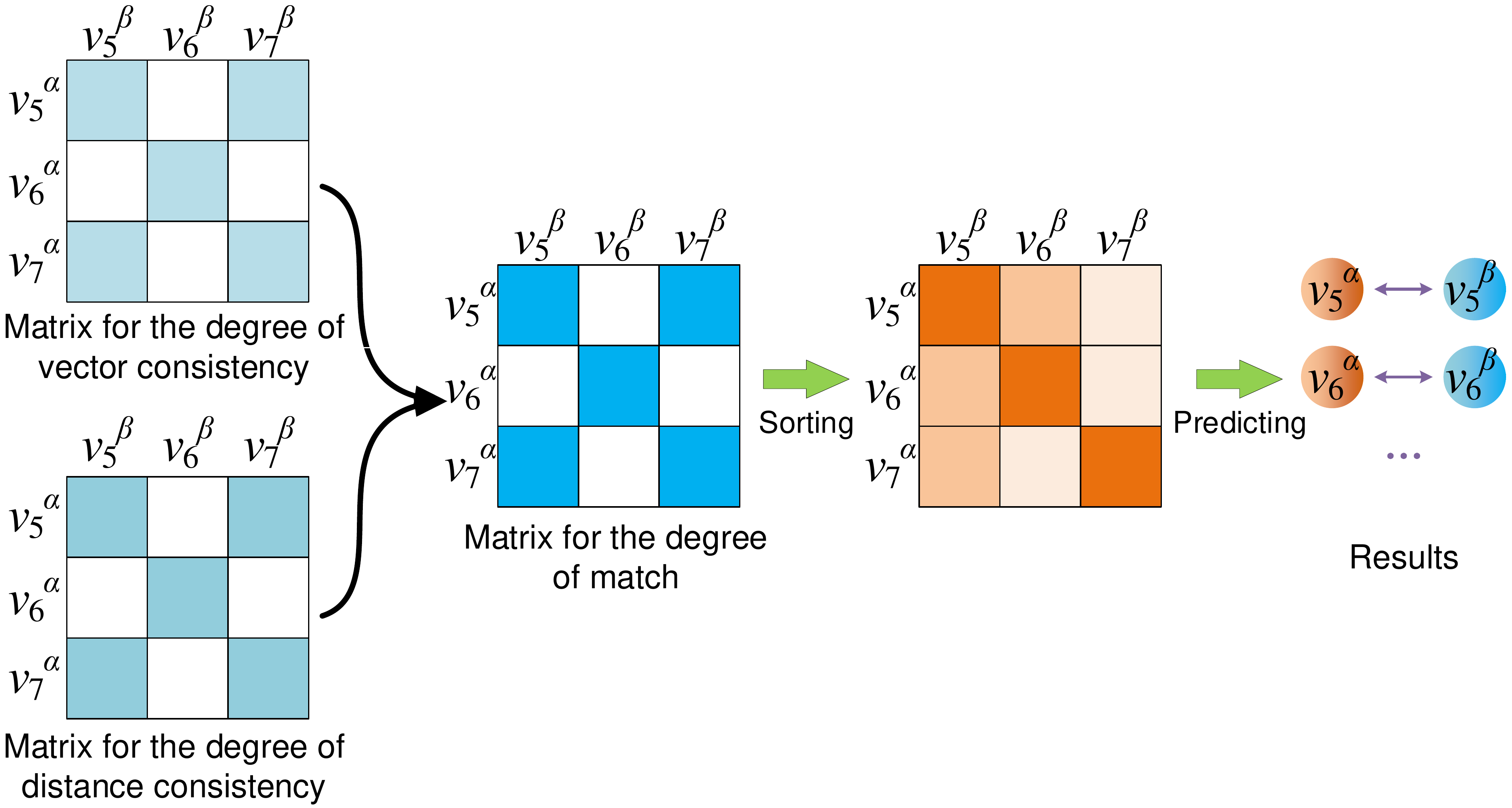}
    \caption{Example of making prediction. By associating the two types of consistency, the matrix for the degree of match can be obtained. Then, we sort the matrix by row. In the third subfigure, the brighter the color, the greater the degree of match. Finally, the prediction results can be provided by the sorted list.}
    \label{pic:prediction}
\end{figure}

The example of different steps of MulCEV are as shown in Figs.~\ref{pic:extension},~\ref{pic:embedding},~\ref{pic:degreeofmatch} and~\ref{pic:prediction}.

\section{Details of Optimization}
To reduce the time complexity, we optimized the calculation of the degrees of vector consistency and distance consistency.
\subsection{Optimization of Vector Consistency Calculation}
The degree of vector consistency for each unmatched node pair can be calculated using Eq.~(10) of the paper. However, many calculations are repeated, such as that of $||\bm{\mathrm{u}}^\beta_b||$. We propose an approach based on a matrix operation to reduce the computational time complexity.

For all unmatched nodes, denoting $\bm B^\alpha =[\bm{\mathrm{u}}^\alpha_1,\bm{\mathrm{u}}^\alpha_2,\dots,\bm{\mathrm{u}}^\alpha_{n^\alpha-n}]$, $\bm B^\beta =[\bm{\mathrm{u}}^\beta_1,\bm{\mathrm{u}}^\beta_2,\dots,\bm{\mathrm{u}}^\beta_{n^\beta-n}]$, $\phi(\bm B^\alpha) =[\phi(\bm{\mathrm{u}}^\alpha_1),\phi(\bm{\mathrm{u}}^\alpha_2),\dots,\phi(\bm{\mathrm{u}}^\alpha_{n^\alpha-n})]$, $\bm b^\alpha=[||\bm{\mathrm{u}}^\alpha_1||,||\bm{\mathrm{u}}^\alpha_2||,\dots,||\bm{\mathrm{u}}^\alpha_{n^\alpha-n}||]^{\mathrm{T}}$, $\bm b^\beta=[||\bm{\mathrm{u}}^\beta_1||,||\bm{\mathrm{u}}^\beta_2||,\dots,||\bm{\mathrm{u}}^\beta_{n^\beta-n}||]^{\mathrm{T}}$, and $\bm \phi( b^\alpha)=[||\phi(\bm{\mathrm{u}}^\alpha_1)||,||\phi(\bm{\mathrm{u}}^\alpha_2)||,\dots,||\phi(\bm{\mathrm{u}}^\alpha_{n^\alpha-n})||]^{\mathrm{T}}$, the degree of vector consistency for all unmatched node pairs can be expressed as
\begin{equation}
\bm{\mathrm{P}}=\frac{\phi(\bm{\mathrm{B}}^\alpha)^{\mathrm{T}} \cdot \bm{\mathrm{B}}^\beta}{\phi(\bm{\mathrm{b}}^\alpha)^{\mathrm{T}} \cdot \bm{\mathrm{b}}^\beta}.
\label{eq:matchdegree_gm_all}
\end{equation}

\subsection{Optimization of Distance Consistency Calculation}
If $h_{ai-bj}$ is denoted by $\mathrm{exp}(-(s^\alpha_{ai} \cdot |s^\alpha_{ai}-s^\beta_{bj}| \cdot s^\beta_{bj}))$, it is clear that if interlayer node pair $(v^\alpha_i,v^\beta_j)$ is the CMN of unmatched node pair $(u^\alpha_a,u^\beta_b)$, $e^\alpha_{ai}$ and $e^\beta_{bj}$ will be equal to 1; thus, $e^\alpha_{ai} \cdot h_{ai-bj} \cdot e^\beta_{bj}=h_{ai-bj}$. In contrast, if interlayer node pair $(v^\alpha_i,v^\beta_j)$ is not the CMN of unmatched node pair $(u^\alpha_a,u^\beta_b)$, $e^\alpha_{ai}$ or $e^\beta_{bj}$ will be equal to 0; thus, $e^\alpha_{ai} \cdot h_{ai-bj} \cdot e^\beta_{bj}=0$. Therefore, Eq.~(11) can be rewritten as
\begin{equation}
q(u^\alpha_a,u^\beta_b)=\sum_{\forall(v^\alpha_i,v^\beta_j)\in \Phi} e^\alpha_{ai} \cdot h_{ai-bj} \cdot e^\beta_{bj}.
\label{eq:matchedegree_dc_1}
\end{equation}

\begin{table*}[!th]
\centering
\caption{Statistics of Real-world Datasets. $|V|$ and $|E|$ are the number of nodes and intralayer links respectively. $k_{max}$ is the maximum degree, $\langle{k}\rangle$ is the average degree, $r$ is the degree-degree correlation, $c$ is the clustering coefficient, $H$ is the degree heterogeneity, as $H=\langle{k^2}\rangle//{\langle{k}\rangle}^2$, and $|E^{\alpha\beta}|$ is the number of interlayer links.}
\label{table}
\setlength{\tabcolsep}{3pt}
\begin{tabular}{ccccccccc}
\hline
Network&$|V|$&$|E|$& $k_{max}$& $\langle{k}\rangle$ & $r$& $c$& $H$&$|E^{\alpha\beta}|$\\
\hline
Foursquare&5,313&76,972&552&20.42&$-0.193$&0.23&3.446&\multirow{2}*{3,148}\\
Twitter&5,120&164,920&1725&51.01&$-0.214$&0.30&4.489&~\\
\hline
DBLP\_DataMining&11,526&47,326&117&36.68&$0.110$&0.85&2.176&\multirow{2}*{1,295}\\
DBLP\_MachineLearning&12,311&43,948&552&20.42&$-0.193$&0.23&3.446&~\\
\hline
Higgs\_FS&4,288&122,826&1365&57.29&$-0.140$&0.27&2.943&\multirow{2}*{3,760}\\
Higgs\_MT&3,777&13,413&1072&7.10&$-0.095$&0.22&9.619&~\\
\hline
Higgs\_FS&4,184&101,618&1086&48.57&$-0.106$&0.27&2.840&\multirow{2}*{3,219}\\
Higgs\_RT&3,238&13,571&626&8.38&$-0.090$&0.09&5.653&~\\
\hline
\end{tabular}
\label{tab:datasets}
\end{table*}

If node $v^\alpha_i$ is an a priori interlayer node in layer $\alpha$, a counterpart node must exist in layer $\beta$, and vice versa. Based on this, we can make the a priori interlayer nodes uniform, as follows: $(v^\alpha_1$, $v^\beta_1),\dots,(v^\alpha_i$, $v^\beta_i),\dots,(v^\alpha_n$, $v^\beta_n)$. Therefore, Eq.~(\ref{eq:matchedegree_dc_1}) can be replaced with
\begin{equation}
q(u^\alpha_a,u^\beta_b)=\sum_{i=1}^{n} e^\alpha_{ai} \cdot h_{ai-bj} \cdot e^\beta_{bi}.
\label{eq:matchedegree_dc_2}
\end{equation}

Using the vector form, Eq.~(\ref{eq:matchedegree_dc_2}) can be replaced by
\begin{equation}
q(u^\alpha_a,u^\beta_b)=[e^\alpha_{a1},\dots,e^\alpha_{ai},\dots,e^\alpha_{an}] \cdot
\left[
 \begin{array}{c}
 h_{a1-b1} \cdot e^\beta_{b1}\\
 \vdots\\
 h_{ai-bi} \cdot e^\beta_{bi}\\
 \vdots\\
h_{an-bn} \cdot e^\beta_{bn}\\
 \end{array}
\right].
\label{eq:matchedegree_dc_3}
\end{equation}

We can use the Hadamard product to rewrite $[h_{a1-b1} \cdot e^\beta_{b1}, \dots, h_{ai-bi} \cdot e^\beta_{bi}, \dots, h_{an-bn} \cdot e^\beta_{bn}]^{\mathrm{T}}$ as $[h_{a1-b1}, \dots,  h_{ai-bi}, \dots, h_{an-bn}]^{\mathrm{T}} \circ [e^\beta_{b1},\dots, e^\beta_{bi}, \dots, e^\beta_{bn}]^{\mathrm{T}}$.
By denoting $\bm{\mathrm{h}}_{ab}=[h_{a1-b1}, \dots,  h_{ai-bi}, \dots, h_{an-bn}]^{\mathrm{T}}$, $\bm{\mathrm{e}}^\alpha_a=[e^\alpha_{a1},\dots,e^\alpha_{ai},\dots,e^\alpha_{an}]^{\mathrm{T}}$, $\bm{\mathrm{e}}^\beta_b=[e^\beta_{b1},\dots, e^\beta_{bi}, \dots, e^\beta_{bn}]^{\mathrm{T}}$, Eq.~(\ref{eq:matchedegree_dc_3}) can be rewritten as
\begin{equation}
q(u^\alpha_a,u^\beta_b)=(\bm{\mathrm{e}}^\alpha_a)^{\mathrm{T}} \cdot (\bm{\mathrm{h}}_{ab} \circ \bm{\mathrm{e}}^\beta_b).
\label{eq:matchedegree_dc_4}
\end{equation}

By using Eq.~(\ref{eq:matchedegree_dc_4}), the degree of distance consistency for unmatched node pair $(u^\alpha_a,u^\beta_b)$ can be represented in vector operation form. Then, if we want to obtain the degree of distance consistency between node $u^\alpha_a$ and all the unmatched nodes in layer $\beta$, we can express Eq.~(\ref{eq:matchedegree_dc_4}) in matrix operation form. In a similar manner, we denote $\bm{\mathrm{H}}=[\bm{\mathrm{h}}_{a1},\dots,\bm{\mathrm{h}}_{ab},\dots,\bm{\mathrm{h}}_{a(n^\beta-n)}]$, $\bm{\mathrm{E}}^\beta=[\bm{\mathrm{e}}^\beta_1,\dots, \bm{\mathrm{e}}^\beta_b, \dots, \bm{\mathrm{e}}^\beta_{n^\beta-n}]$. The degree of distance consistency between node $u^\alpha_a$ and all the unmatched nodes in layer $\beta$ can be calculated as follows:
\begin{equation}
\bm{\mathrm{q}}^\alpha_a=((\bm{\mathrm{e}}^\alpha_a)^{\mathrm{T}} \cdot (\bm{\mathrm{H}} \circ \bm{\mathrm{E}}^\beta))^{\mathrm{T}}.
\label{eq:matchedegree_dc_5}
\end{equation}

By denoting $\bm{\mathrm{s}}^\alpha_{a}=[s^\alpha_{a1},\dots,s^\alpha_{ai},\dots,s^\alpha_{an}]$, $\bm{\mathrm{s}}^\beta_{b}=[s^\beta_{b1},\dots,s^\beta_{bi},\dots,s^\beta_{bn}]^{\mathrm{T}}$, $\bm{\mathrm{S}}^\beta=[\bm{\mathrm{s}}^\beta_{1},\dots,\bm{\mathrm{s}}^\beta_{b},\dots,\bm{\mathrm{s}}^\beta_{n^\beta-n}]^{\mathrm{T}}$, matrix $\bm{\mathrm{H}}$ can be calculated as
\begin{equation}
\bm{\mathrm{H}}=\mathrm{exp}\{(\bm{\mathrm{i}} \cdot \bm{\mathrm{s}}^\alpha_{a}) \circ |\bm{\mathrm{i}} \cdot \bm{\mathrm{s}}^\alpha_{a}-\bm{\mathrm{S}}^\beta| \circ \bm{\mathrm{S}}^\beta \},
\end{equation}
where $\bm{\mathrm{i}}$ is a column vector with $n^\beta-n$ elements. The value of each element in vector $\bm{\mathrm{i}}$ is 1.

By joining the vector for the degree of distance consistency for all unmatched nodes in layer $\alpha$, we can obtain the matrix for the degree of distance consistency for all unmatched node pairs, which can be represented as $\bm {\mathrm{Q}}=[\bm{\mathrm{q}}^\alpha_1,\dots,\bm{\mathrm{q}}^\alpha_a,\dots,\bm{\mathrm{q}}^\alpha_{n^\alpha-n}]^{\mathrm{T}}$.

Finally, the matrix for the degree of match for all unmatched node pairs can be obtained by
\begin{equation}
\bm {\mathrm{R}}=\delta \cdot \bm {\mathrm{P}}+(1-\delta) \cdot \bm {\mathrm{Q}}.
\end{equation}
 The interlayer link prediction results are obtained by ranking each row or column of $\bm {\mathrm{R}}$ in reverse order according to the degree of match.

\section{Time Complexity Analysis}
In the cross-layer extension stage, we search the intralayer links from the adjacency matrix comprising the matched nodes such that the time complexity of this stage is $O(n^2)$. In the network embedding stage, the time complexity for weight calculation is $O(|\Phi|\langle k\rangle^2)$, where $|\Phi|$ is the number of a priori intralayer links and $\langle k\rangle$ is the average degree of the nodes. Meanwhile, the time complexity for network embedding by LINE is $O(d|\Phi|\iota)$, where $\iota$ is the number of negative samples~\cite{TangJian2015}. Therefore, the total time complexity in this step is $O(|\Phi|(d\iota+\langle k\rangle^2))$. In the degree of match calculation stage, the time complexity for training the mapping function, MLP, is $O(kdn)$~\cite{ManTong2016-IJCAI}. The time complexity of for calculating the vector consistency for all unmatched node pairs is $O(d(n^\alpha-n)(n^\alpha-n)(n^\beta-n))$ and that for calculating the distance consistency for all unmatched node pairs is $O(dn(n^\alpha-n)(n^\beta-n))$. We use matrix multiplication to optimize the process of calculating the degree of match for all the unmatched node pairs. After optimization, the time complexity for calculating the two types ofconsistency are $O(d(n^\alpha-n)(n^\alpha-n)(n^\beta-n)/\varsigma)$ and $O(dn(n^\alpha-n)(n^\beta-n)/\varsigma)$, respectively, where $\varsigma$ is the number of computational nodes~\cite{lee1997IO}. Suppose the number of unmatched nodes in each layer is $n_u$, i.e., $n_u=(n^\alpha-n)=(n^\beta-n)$; then, the total time complexity for calculating the degree of match is $O(kdn+n_u^3d/\varsigma+nn_u^2d/\varsigma)$. Lastly, the time complexity for predicting the interlayer links is $O(Nn_u^2)$, where $N$ is the size of top-$N$ list.

\section{Details of Experimental Configurations}
\subsection{Datasets}
To evaluate the performance of our proposed framework and baseline methods, we used three synthetic and four real-world multiplex network datasets in our experiments. The synthetic networks are Erd\H{o}s-R\'{e}nyi~\cite{erdHos1960evolution} (ER) random networks, Watts-Strogatz~\cite{watts1998collective} (WS) small-world networks, and Barab\'{a}si-Albert~\cite{Barabasi1999-BA} (BA) networks.
We leveraged method proposed in Ref.~\cite{tang2020interlayer} to generate multiplex networks by these three synthetic networks. We set the network size to 2000 and the percentage to remaining nodes to 0.5. The real-world datasets are as follows (cf. Table~\ref{tab:datasets}):
\begin{itemize}
\item \textbf{Foursquare--Twitter (FT)}: This dataset was collected from Foursquare and Twitter by Zhang et al.~\cite{ZhangJiawei2015-IJCAI}. The ground truth for this dataset is provided in Foursquare's profiles, and the nodes of the two social networks are partially aligned.
\item \textbf{DBLP\_DataMining-DBLP\_MachineLearning (DBLP)}: This dataset was collected from the Citation Network Dataset\footnotemark[3] \footnotetext[3]{https://www.aminer.cn/citation}~\cite{tang2008extraction} and processed by Liu et al.~\cite{liu2019structural}. It is a co-authored multiplex network, one layer of which consists of researchers who published articles in journals or conference proceedings related to data mining, and the other layer containing researchers who published articles in journals or conference proceedings related to machine learning. The ground truth was obtained by collecting the authors who published articles in both fields.
\item \textbf{Higgs\_Friendships-Higgs\_Mention (Higgs-FSMT)}: The Higgs dataset is collected from Twitter by Domenico et al.~\cite{de2013anatomy} which focuses on the spreading processes of the messages on Twitter during and after the discovery of a new particle with the features of the Higgs boson. We choose the friendships (FS) and mention (MT) networks to construct the multiplex network. To facilitate processing, we only reserve nodes with degrees greater than five in each network.
\item \textbf{Higgs\_Friendships-Higgs\_Retweet (Higgs-FSRT)}: We choose the FS and retweet (RT) networks of Higgs to construct the multiplex network.
\end{itemize}

\subsection{Comparison Methods}
\begin{table*}[htb]
\centering
\caption{Performance of MulCEV on different $\delta$. }
\label{table}
\setlength{\tabcolsep}{3pt}
\begin{tabular}{cccccccccccccc}
\hline
\multirow{2}*{\textbf{Metric}} &\multirow{2}*{\textbf{Datasets}} & \multirow{2}*{\textbf{Training ratios}}
& \multicolumn{11}{c}{$\delta$} \\ \cline{4-14}
&&& 0.0& 0.1& 0.2& 0.3& 0.4& 0.5& 0.6& 0.7& 0.8& 0.9& 1.0\\
\hline
\multirow{21}*{$\bm{P@30}$}
&\multirow{3}*{FT}
&0.3 &0.4622&0.4822&0.4891&0.4974&0.5048&0.5153&0.5327&0.5509&0.5636&\textbf{0.5640}&0.5001\\
&&0.6 &0.5828&0.5884&0.5941&0.5982&0.6071&0.6169&0.6331&0.6534&0.6753&\textbf{0.6802}&0.6047\\
&&0.9 &0.6645&0.6738&0.6738&0.6801&0.6738&0.6957&0.7012&0.7031&0.7112&\textbf{0.7174}&0.7051\\
\cline{2-14}
&\multirow{3}*{DBLP}
&0.3 &0.2407& 0.2413& 0.2416& 0.2418& \textbf{0.2419}& 0.2418& 0.2366& 0.2255& 0.2143& 0.2043& 0.1726\\
&&0.6 &0.3782& 0.3796& 0.3796& 0.3796& 0.3796& \textbf{0.3809}& 0.3649& 0.3328& 0.3087& 0.2900& 0.2272\\
&&0.9 &0.5152& 0.5152& 0.5152& 0.5152& \textbf{0.5291}& 0.5152& 0.4943& 0.4456& 0.3759& 0.2785& 0.1601\\
\cline{2-14}
&\multirow{3}*{Higgs-FSMT}
&0.3&0.4310&0.4402&0.4431&0.4435&0.4442&0.4472&\textbf{0.4538}&0.4505&0.4417&0.4204&0.3621\\
&&0.6 &0.6351&0.6430&0.6476&0.6509&0.6555&\textbf{0.6691}&0.6686&0.6647&0.6364&0.6054&0.5092\\
&&0.9 &0.7909&0.7936&0.7989&0.8016&0.8043&\textbf{0.8170}&0.8097&0.7882&0.7614&0.7239&0.6434\\
\cline{2-14}
&\multirow{3}*{Higgs-FSRT}
&0.3&0.4996&0.5139&0.5175&0.5261&0.5306&\textbf{0.5496}&0.5459&0.5477&0.5387&0.5148&0.4573\\
&&0.6 &0.6924&0.7003&0.7019&0.7082&0.7098&\textbf{0.7261}&0.7224&0.7192&0.7177&0.6822&0.6065\\
&&0.9 &0.8317&0.8447&0.8479&0.8511&0.8511&0.8511&\textbf{0.8608}&\textbf{0.8608}&0.8544&0.8091&0.7314\\
\cline{2-14}
&\multirow{3}*{ER}
&0.3&\textbf{0.8994}&\textbf{0.8994}&\textbf{0.8994}&\textbf{0.8994}&\textbf{0.8994}&\textbf{0.8994}&\textbf{0.8994}&\textbf{0.8994}&0.8609&0.7633&0.6331\\
&&0.6 &\textbf{0.9797}&\textbf{0.9797}&\textbf{0.9797}&\textbf{0.9797}&\textbf{0.9797}&\textbf{0.9797}&\textbf{0.9797}&0.9746&0.9594&0.9086&0.7614\\
&&0.9 &\textbf{0.9943}&\textbf{0.9943}&\textbf{0.9943}&\textbf{0.9943}&\textbf{0.9943}&\textbf{0.9943}&\textbf{0.9943}&0.9924&0.9880&0.9627&0.8335\\
\cline{2-14}
&\multirow{3}*{WS}
&0.3&\textbf{0.9117}&\textbf{0.9117}&\textbf{0.9117}&\textbf{0.9117}&\textbf{0.9117}&\textbf{0.9117}&\textbf{0.9117}&0.9088&0.9031&0.886&0.8319\\
&&0.6 &\textbf{0.9853}&\textbf{0.9853}&\textbf{0.9853}&\textbf{0.9853}&\textbf{0.9853}&\textbf{0.9853}&\textbf{0.9853}&\textbf{0.9853}&\textbf{0.9853}&0.9706&0.9363\\
&&0.9 &\textbf{0.9980}&\textbf{0.9980}&\textbf{0.9980}&\textbf{0.9980}&\textbf{0.9980}&\textbf{0.9980}&\textbf{0.9980}&\textbf{0.9980}&\textbf{0.9980}&0.9899&0.9446\\
\cline{2-14}
&\multirow{3}*{BA}
&0.3&0.8049&0.8074&0.8071&0.8099&0.8113&\textbf{0.8125}&0.8116&0.8110&0.7948&0.7554&0.6720\\
&&0.6 &0.9269&0.9269&0.9259&0.9279&0.9279&\textbf{0.9294}&0.9284&0.9269&0.9141&0.8847&0.7958\\
&&0.9 &0.9681&0.9681&0.9681&0.9721&0.9723&\textbf{0.9781}&0.9681&0.9681&0.9620&0.9370&0.8567\\
\hline
\end{tabular}
\label{tab:choose_delta}
\end{table*}

\begin{table*}[htb]
\centering
\caption{Supplementary experiments of different $\delta$ on ER and WS datasets. }
\label{table}
\setlength{\tabcolsep}{3pt}
\begin{tabular}{cccccccccccccc}
\hline
\multirow{2}*{\textbf{Metric}} &\multirow{2}*{\textbf{Datasets}} & \multirow{2}*{\textbf{Training ratios}}
& \multicolumn{11}{c}{$\delta$} \\ \cline{4-14}
&&& 0.0& 0.1& 0.2& 0.3& 0.4& 0.5& 0.6& 0.7& 0.8& 0.9& 1.0\\
\hline
\multirow{6}*{$\bm{P@1}$}
&\multirow{3}*{ER}
&0.3 &0.4909&0.5188&0.5349&0.5577&0.5695&0.5725&\textbf{0.5761}&0.5746&0.5291&0.4110&0.1725\\
&&0.6 &0.8274&0.8376&0.8477&0.8528&0.8528&\textbf{0.8681}&0.8630&0.8630&0.8274&0.6954&0.2843\\
&&0.9 &0.9621&0.9688&0.9643&0.9680&0.9680&\textbf{0.9696}&0.9508&0.9231&0.9148&0.8582&0.3632\\
\cline{2-14}
&\multirow{3}*{WS}
&0.3 &0.5499&0.5755&0.584&0.5954&0.6182&0.6282&\textbf{0.6296}&0.6239&0.6154&0.5299&0.2877\\
&&0.6 &0.8382&0.8382&0.8431&0.8676&0.8676&0.8873&\textbf{0.8971}&\textbf{0.8971}&0.8775&0.7892&0.4461\\
&&0.9 &0.9508&0.9512&0.9491&0.9484&0.9495&\textbf{0.9534}&0.9522&0.9529&0.9380&0.8891&0.5486\\
\cline{2-14}
\hline
\end{tabular}
\label{tab:choose_delta2}
\end{table*}
We used several state-of-the-arts as baselines, which are as follows.
\begin{itemize}
\item \textbf{DeepLink}: DeepLink is a semi-supervised learning algorithm that leverages traditional random walks to generate social sequences for the network embedding and utilizes the duality of mapping to improve the prediction performance.
\item \textbf{IONE}: Input--output network embedding (IONE) projects multiple social networks into a common embedded space and matches same-user accounts by calculating the cosine similarity between the vectors of two nodes. In IONE, it represents each account by three vectors: a node vector, an input context vector, and an output context vector.
\item \textbf{ONE}: This method is a simplified version of IONE. In this method, an account is represented by two vectors: a node vector and an output context vector.
\item \textbf{IONE-D}: This method is a refined version of IONE that explores the community structure of the SMNs and incorporates the structural diversity to characterize a set of interlayer links.
\item \textbf{BootEA}: This is a bootstrapping approach that aligns the entities of different knowledge graphs based on network embedding. It iteratively labels potential entity pairs as training data to overcome the lack of a sufficiently large training set and leverages an editing method to reduce error accumulation during the iterations.
\item \textbf{PALE}: This method projects each SMN into a unique low-dimensional space and represents nodes by low-dimensional vectors in a latent space. Then, it learns a cross-layer mapping function for predicting interlayer links.
\item \textbf{MAH}: Manifold alignment on hypergraph (MAH) tries to map common users across SMNs based on the network embedding method. It adopts a hypergraph to model high-order relations of SMNs and represents nodes into a common latent space. It infers correspondence by comparing distances between the vectors of the unmatched nodes.
\item \textbf{MAG}: Manifold alignment on traditional graphs (MAG) is a method that uses $w(u_i,u_j)=|R_{u_i}\cap R_{u_j}|/(|R_{u_i}|+R_{u_j})$ for the calculation of node-to-node pairwise weights to build a graph for each SMN. The method for obtaining the node ranking result is the same as that for MAH.
\item \textbf{CRW}: Collective random walk (CRW) is a joint link fusion approach for predicting the intralayer links and interlayer links simultaneously; it transfers information relating to intralayer links from one layer to another.
\end{itemize}

\subsection{The Other Experimental Configurations}
We employed $Precision@N$ ($P@N$)~\cite{LiuLi2016,ShuKai2017}, F-measure ($F1$)~\cite{ShuKai2017}, and $MAP$~\cite{ShuKai2017} as the metrics to evaluate the performance of all methods. $P@N$ is defined as
\begin{equation}
P@N=\sum_{i=1}^m{\mathds{1}_i\{success@N\}/m},
\end{equation}
where ${\mathds{1}_i\{success@N\}}$ indicates whether the correct interlayer link exists in the top-$N$ list, and $m$ represents the number of all unobserved interlayer links. It is noteworthy that $Precision@N$ is actually the same as $Recall@N$ and $F1@N$ in the field of interlayer link prediction because $Precision@N$ represents the true positive prediction rate. 

$MAP$ is used to evaluate the ranked performance of different methods and is defined as
\begin{equation}
MAP=(\sum_{i=1}^n{\frac{1}{r_i}})/m,
\end{equation}
where $r_i$ represents the rank of the $i$th unmatched interlayer link. The higher the values of $P@N$ and $MAP$, the better the performance of the method.

To test the performance, the set of all interlayer links was randomly divided into two parts for each experiment (i) a training set $\Phi$, which was treated as the set of a priori interlayer links; and (ii) a test set $\Psi$, which was used for testing and can be considered a collection of the unmatched node pairs waiting for prediction. The ratio of the size of the training set to the size of the set of all interlayer links is called the training ratio, which we varied in some of the experiments. Our task was to uncover the interlayer links in the test set based on the information in the training set and each layer of the multiplex network. For the experiments with a training ratio of $90\%$, we adopted 10-fold cross-validation. For the other experiments, we conducted ten times with randomly divided training and testing sets and took the average values as the results.

\section{Details of Effect of Control Parameter $\delta$}
In Eq.~(12) of the paper, the parameter $\delta$ is leveraged to control the proportions of the vector consistency and distance consistency in the final degree of match. We studied the initialization strategy for $\delta$ and its effect on the predicted results through experiments. We set $30.0\%, 60.0\%$, and $90.0\%$ of the interlayer links as the training set and the remaining links as the test set. The value of $\delta$ was varied from 0 to 1 in steps of 0.1.

The $P@30$ of these experiments are displayed in Table~\ref{tab:choose_delta}. From the table, we see that the values of $P@30$ initially exhibit an increasing trend and later decrease with additional increase in $\delta$. This demonstrates that both the vector consistency and distance consistency positively affect interlayer link predictions. Maximum values in each line are obtained for different $\delta$ values; this may be attributed to the difference in network structural properties between these datasets.
For some datasets, the mapping functions can be learned well; hence, a small $\delta$ achieves the best prediction results. For other datasets, the mapping functions are hard to learn; these maximum values are considered for larger $\delta$. We observed that the maximum values were achieved for multiple columns in the ER and WS datasets. This renders it difficult to choose an appropriate value of $\delta$. To overcome this problem, we further tested $P@1$ on the two datasets, as shown in Table~\ref{tab:choose_delta2}. The results exhibit the same trends as the other datasets at $P@30$.

Considering that the prediction results of most datasets show best performance with $\delta=0.5$, this value is recommended for new datasets. For more precise values, investigators may consider performing 10-fold cross validation using the a priori interlayer link set, which is divided into training and validation sets.

\section{Credit Authorship Contribution Statement}
Rui Tang: Conceptualization, Methodology, Software, Validation, Investigation, Writing - original draft, Writing - review $\&$ editing. Xingshu Chen: Conceptualization, Methodology, Project administration, Resources, Writing - review $\&$ editing. Haizhou Wang: Conceptualization, Methodology, Project administration, Writing - review $\&$ editing. Zhenxiong Miao: Writing - review $\&$ editing, Validation, Investigation. Shuyu Jiang: Writing - review $\&$ editing, Supervision. Wei Wang: Supervision, Visualization, Investigation.

\section{Acknowledgment}
The authors want to thank Dr. Wenxian Wang and Mingyong Yin for their advice.

\end{appendices}

\end{document}